\documentclass[5p,authoryear]{elsarticle} 
\usepackage[utf8]{inputenc}
\usepackage{graphicx}
\usepackage{framed}
\usepackage{amsfonts} 
\usepackage{amssymb}
\usepackage{amsmath}
\usepackage{mathtools}
\usepackage{amsthm}
\usepackage{subfigure}
\usepackage{multirow}
\usepackage{url}
\usepackage{lineno}
\usepackage{setspace} 

\journal{Pattern Recognition Letters}

\DeclareFontFamily{U}{mathb}{\hyphenchar\font45}
\DeclareFontShape{U}{mathb}{m}{n}{
      <5> <6> <7> <8> <9> <10> gen * mathb
      <10.95> mathb10 <12> <14.4> <17.28> <20.74> <24.88> mathb12
      }{}
\DeclareSymbolFont{mathb}{U}{mathb}{m}{n}
\DeclareFontSubstitution{U}{mathb}{m}{n}

\DeclareMathSymbol{\uptodownarrow}         {3}{mathb}{"FE}
\DeclareMathSymbol{\downtouparrow}         {3}{mathb}{"FF}

\newcommand{\adjoint}[1][\,]{\ensuremath{^{A\protect{#1}}}}
\newcommand{\C}[1][\,]{\ensuremath{\protect{\mathbb{C}_{\protect{#1}}}}}
\newcommand{\G}{\ensuremath{\mathbb{G}}}
\newcommand{\N}{\ensuremath{\protect{\mathbb{N}}}}
\newcommand{\K}{\ensuremath{\protect{\mathbb{K}}}}
\newcommand{\PC}[1][\,]{\ensuremath{\protect{\mathcal{P}(\C[#1])}}}

\renewcommand{\L}[1][]{\ensuremath{\protect{\mathcal{L}_{#1}}}}
\newcommand{\subcomplexes}[1][]{\ensuremath{\mathcal{C}_{\protect{#1}}}}
\newcommand{\stars}[1][]{\ensuremath{\mathcal{S}_{\protect{#1}}}}
\newcommand{\DomIm}[2]{\ensuremath{#1 \rightarrow #2}}

\newcommand{\DomImL}[2]{\ensuremath{\DomIm{\mathcal{L}_{#1}}{\mathcal{L}_{#2}}}}

\newcommand{\dimension}[2]{\ensuremath{#1_{#2}}}

\renewcommand{\iff}{\ensuremath{\protect{\Leftrightarrow}}}

\newcommand{\Dim}[2]{\dimension{#1}{#2}}
\newcommand{\ov}[1]{\overline{#1}}

\newcommand{\dilation}[1][]{\ensuremath{\protect{\delta^{#1}}}}
\newcommand{\erosion}[1][]{\ensuremath{\protect{\varepsilon^{#1}}}}

\newcommand{\dij}{\ensuremath{\delta^{+}_{i,j}}}
\newcommand{\dji}{\ensuremath{\delta^{-}_{j,i}}}

\newcommand{\dilp}[2]{\ensuremath{\delta^{+}_{#1,#2}}}
\newcommand{\dilm}[2]{\ensuremath{\delta^{-}_{#1,#2}}}

\newcommand{\eij}{\ensuremath{\varepsilon^{+}_{i,j}}}
\newcommand{\eji}{\ensuremath{\varepsilon^{-}_{j,i}}}

\newcommand{\erop}[2]{\ensuremath{\varepsilon^{+}_{#1,#2}}}
\newcommand{\erom}[2]{\ensuremath{\varepsilon^{-}_{#1,#2}}}

\DeclareMathOperator*{\asf}{ASF}

\newcommand{\asfC}[1][]{\ensuremath{\asf^{c\protect{#1}}}}

\newcommand{\dilG}{\ensuremath{\delta^{\uptodownarrow}}} 
\newcommand{\eroG}{\ensuremath{\varepsilon^{\uptodownarrow}}}

\newcommand{\opG}[1][]{\ensuremath{\gamma^{\uptodownarrow}_{#1}}}
\newcommand{\clG}[1][]{\ensuremath{\phi^{\uptodownarrow}_{#1}}}

\newcommand{\asfG}[1][]{\ensuremath{\asf^{\uptodownarrow\protect{#1}}}}

\newcommand{\dilS}{\ensuremath{\delta^{\downtouparrow}}} 
\newcommand{\eroS}{\ensuremath{\varepsilon^{\downtouparrow}}}

\newcommand{\paren}[1]{\ensuremath{\left( {#1} \right)}}

\newtheorem{lemma}{Lemma}
\newtheorem{proper}[lemma]{Property}

\newtheorem{defn}[lemma]{Definition}

\usepackage{color,colortbl} 
\newcommand{\ELIMINE}[1]{}
\definecolor{Red}{rgb}{1,0,0} 
\definecolor{Blue}{rgb}{0,0,1}
\newcommand{\st}[0]{\; | \;}

\newcommand{\eg}[0]{{\em e.g.}}

\newcommand{\etal}[0]{\emph{et al. }}

\begin{document}

\begin{frontmatter}
\title{Dimensional operators for mathematical morphology on simplicial complexes} 

\author[esiee,fef]{F.~Dias\corref{cor1}}
\ead{fabio.dias@gmail.com}

\author[esiee]{J.~Cousty}
\ead{j.cousty@esiee.fr}

\author[esiee]{L.~Najman}
\ead{l.najman@esiee.fr}

\address[esiee]{Université Paris-Est, Laboratoire d’Informatique Gaspard-Monge, Equipe A3SI, ESIEE, Paris, France}
\address[fef]{College of Physical Education, State University of Campinas, Brazil}
\cortext[cor1]{Corresponding author}

\begin{abstract}
In this work we study the framework of mathematical morphology on simplicial complex spaces.

Simplicial complexes are widely used to represent multidimensional data, such as meshes, that are two dimensional complexes, or graphs, that can be interpreted as one dimensional complexes.

Mathematical morphology is one of the most powerful frameworks for image processing, including the processing of digital structures, and is heavily used for many applications. However, mathematical morphology operators on simplicial complex spaces is not a concept fully developed in the literature.

Specifically, we explore properties of the dimensional operators, small, versatile operators that can be used to define new operators on simplicial complexes, while maintaining properties from mathematical morphology. These operators can also be used to recover many morphological operators from the literature.  Matlab code and additional material, including the proofs of the original properties, are freely available
at~\url{https://code.google.com/p/math-morpho-simplicial-complexes.}
\end{abstract}

\end{frontmatter}

\section{Introduction and related work}
\label{sec:related}

Simplicial complexes were first introduced by Poincaré in 1895~\citep{Poincare1895} to study the topology of spaces of arbitrary dimension, and are basic tools for algebraic topology~\citep{Maunder96}, image analysis~\citep{Ber07, Couprie2009, Kon97} and discrete surfaces~\citep{Eva96, EKM96, daragon-2005}, among many other domains.

In the form of meshes they are widely used in many contexts to express tridimensional data. Some graphs can be represented as a form of simplicial complexes, and we can build simplicial complexes based on regular, matricial, images. This versatility is the reason we chose to use simplicial complexes as the operating space.

Considering operators on simplicial complex spaces, it is fairly common to change the complexity of the mesh structure~\citep{Chiang2011, DeFloriani1999}. Even when additional data is associated with the elements of the complex, they are mostly used to guide the change in the structure, the values themselves are not changed. Here, we pursuit a different option, our objective is to filter values associated to the elements of the complex, without changing its structure, using the framework of mathematical morphology.

Mathematical morphology was introduced by Matheron and Serra in 1964 and it is one of the most important frameworks for non-linear image processing, providing tools for many applications. It was later extended by Heijmans and Ronse~\citep{Heijmans1990} using complete lattices, allowing the use of more complex digital structures, such as graphs~\citep{Cousty2009, Cousty2013,Ta2008,Book-Lezoray-Grady-2012,Vincent1989}, hypergraphs~\citep{Bloch2011,Bloch2013,Stell10} and simplicial complexes~\citep{Dias2011, Dias2012, Lomenie2008}.

The use of a digital structure as support to image processing is not new. In~\cite{Vincent1989}, Vincent uses the lattice approach to mathematical morphology to define morphological operators on neighborhood graphs, where the graph structure is used to define neighborhood relationships between unorganized data, expressed as vertices.

By allowing the propagation of values from vertices to the edges, therefore using the graph structure to express more than just neighborhood relation, Cousty~\etal~\citep{Cousty2013} obtained different morphological operators, including openings, closings and alternating sequential filters. Those operators are capable of dealing with smaller noise structures, acting in a smaller size than the classical operators. Similarly, Meyer and Stawiaski~\citep{Meyer2009} and Meyer and Angulo~\citep{Meyer2007} obtain a new approach to image segmentation and levellings, respectively.

Recently, Bloch and Bretto~\citep{Bloch2011,Bloch2013} introduced mathematical morphology on hypergraphs, defining lattices and operators. Their lattices and operators are similar to the ones presented here, taking into account the differences between hypergraphs and complexes.

This work is focused on mathematical morphology on simplicial complexes, specifically to process values associated to elements of
the complex in an unified manner, without altering the structure itself. In~\cite{Lomenie2008}, Loménie and Stamon explore mathematical morphology operators on mesh spaces from point spaces. However, the complex only provides structural information, while the information itself is associated only to triangles or edges of the mesh.

Our approach for mathematical morphology on simplicial complexes has been studied before and this article is an extension of the conference article \citep{Dias2011}, where interesting new operators were introduced. These operators, called~\emph{dimensional operators} can be used as building blocks for new operators. In this work we explore these operators, introducing composition properties and defining new morphological operators. The proofs are omitted here, but they are available in~\cite{Dias2012}. We also revisit the related work, showing that most of the operators from the literature can be expressed by the dimensional operators.

\section{Basic theoretical concepts}
\label{ch:basic}
The objective of this work is to explore the dimensional operators for mathematical morphology on simplicial complex spaces. To this end, we start by reminding useful definitions about simplicial complexes and mathematical morphology.
\subsection{Simplicial complexes}
\label{sec:sc}

One of the most known forms of complex is the concept of \emph{mesh}, often used to express tridimensional data on various domains, such as computer aided design, animation and computer graphics in general.  However, in this work we prefer to approach complexes by the combinatorial definition of an abstract complex~\citep{Jaenich1984}.

The basic element of a complex is a \emph{simplex}. In this work, a simplex is a finite, nonempty set.  The \emph{dimension} of a simplex~$x$, denoted by~$dim(x)$, is the number of its elements minus one. A simplex of dimension $n$ is also called an \emph{n-simplex}. We call \emph{simplicial complex}, or simply \emph{complex}, any set $X$ of simplices such that, for any $x\in X$, any non-empty subset of $x$ also belongs to $X$. The \emph{dimension} of a complex is equal to the greatest dimension of its simplices and, by convention, we set the \emph{dimension} of the empty set to $-1$. In the following, a complex of dimension $n$ is also called an \emph{$n$-complex}.

\begin{figure}
  \centering \subfigure[]  {
    \includegraphics[width=0.1\textwidth]{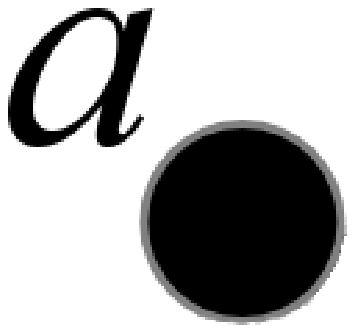}
    \label{figSimA}
  } \subfigure[] {
    \includegraphics[width=0.1\textwidth]{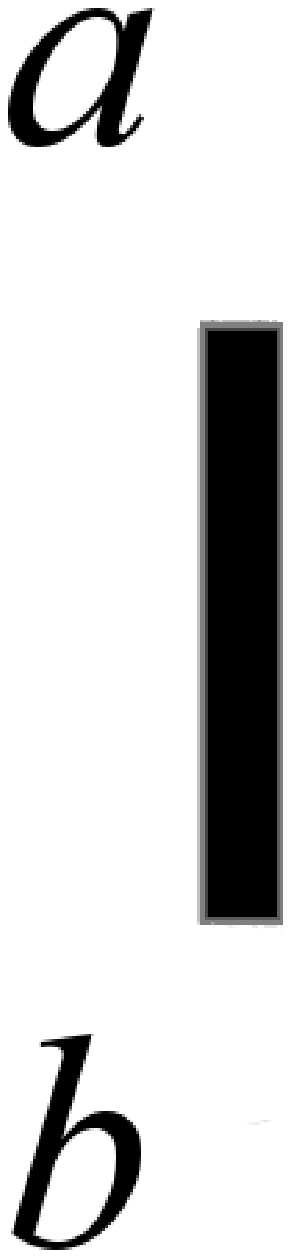}
    \label{figSimB}
  } \subfigure[] {
    \includegraphics[width=0.1\textwidth]{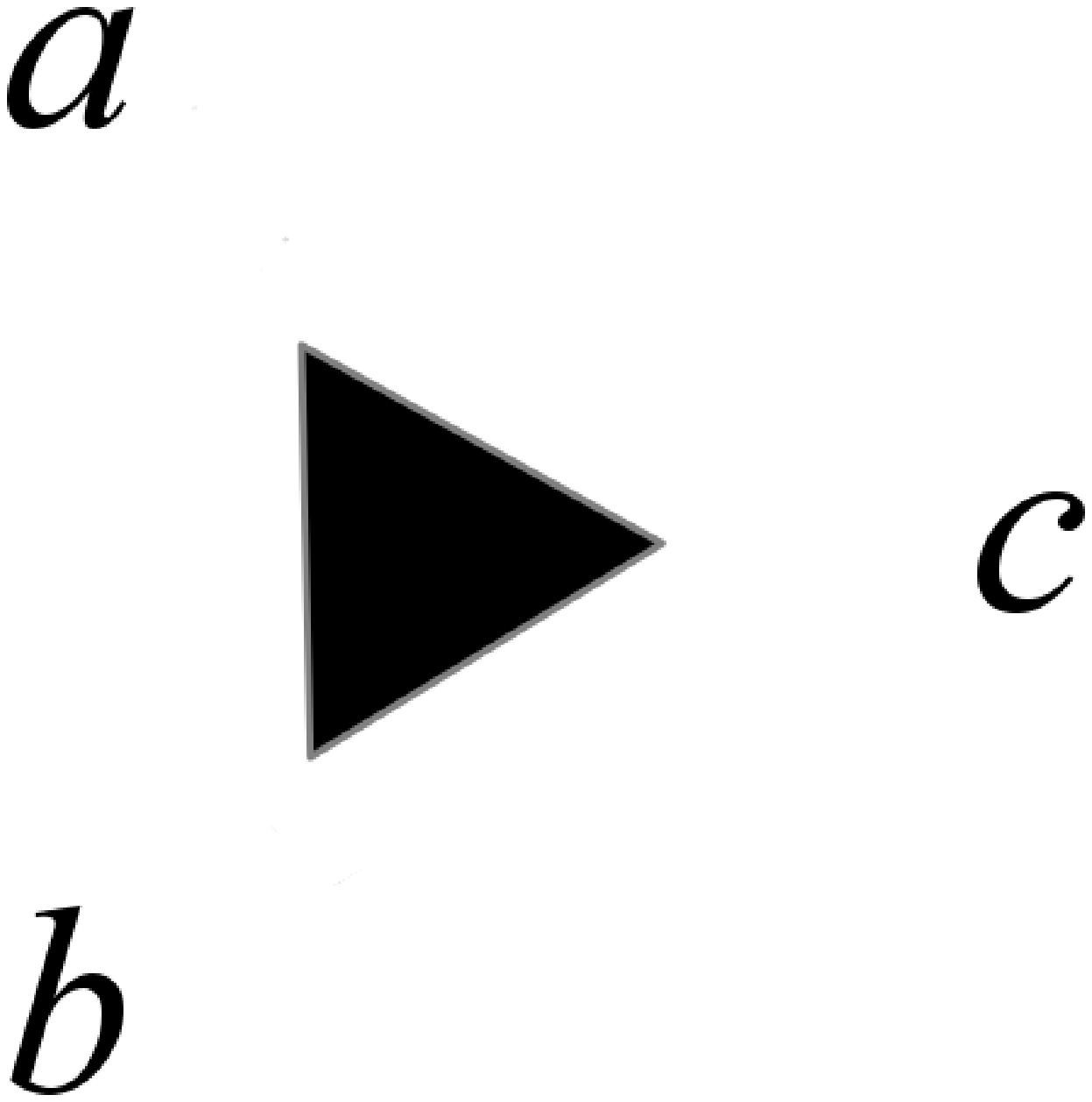}
    \label{figSimC}
  }\subfigure[]  {
    \includegraphics[width=0.1\textwidth]{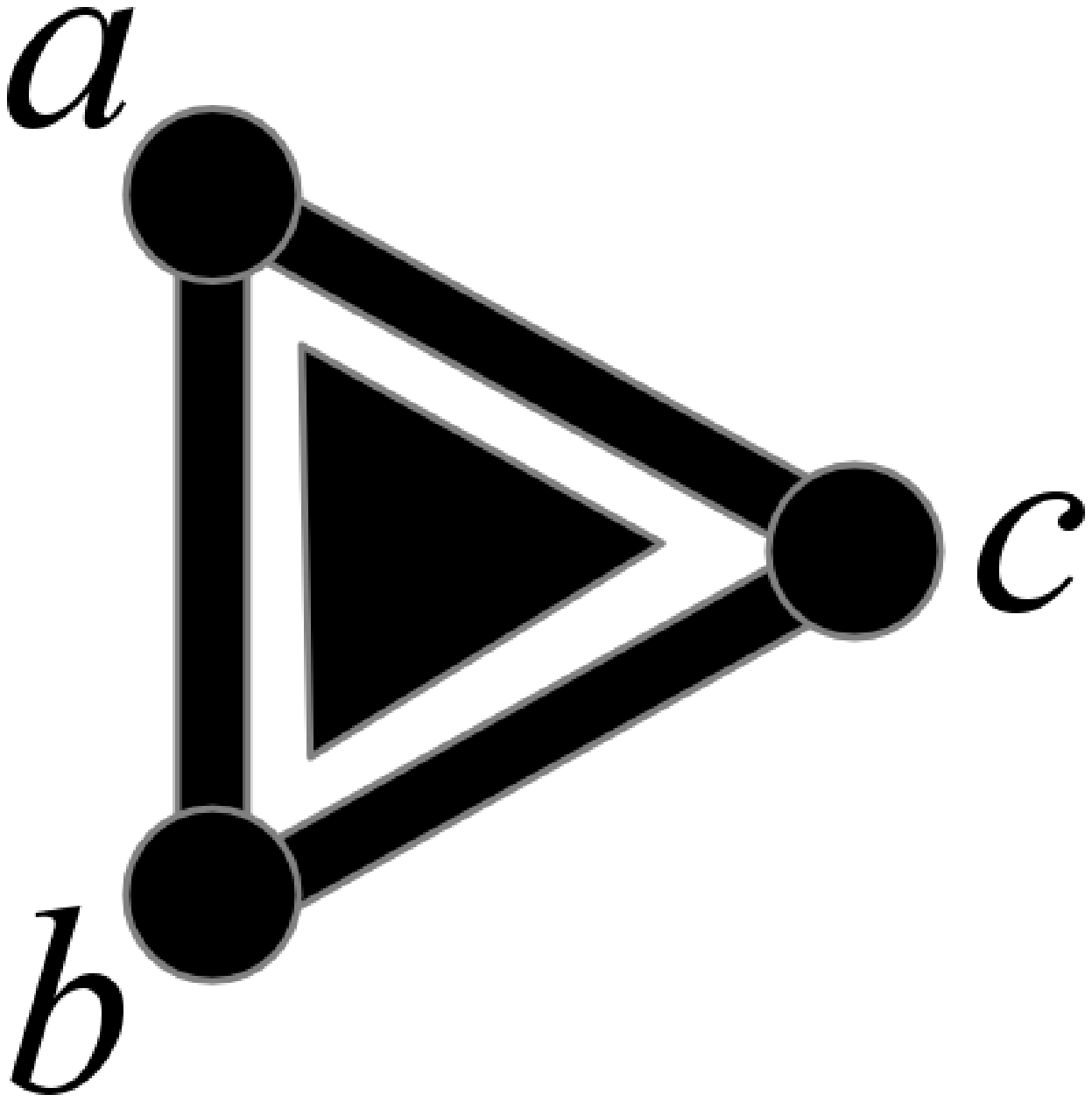}
    \label{figSimD}
  }
  \caption{Graphical representation of {\bf(a)} a $0$-simplex, {\bf(b)} a $1$-simplex, {\bf(c)} a $2$-simplex, and {\bf(d)} a small $2$-complex.}
  \label{figSimplices}
\end{figure}

Figure~\ref{figSimA} (resp. b, and c) graphically represents a simplex~$x = \{a\}$ (resp. $y = \{a,b\}$ and~$z = \{a,b,c\}$) of dimension~$0$ (resp. $1$, $2$). Figure~\ref{figSimD} shows a set of simplices composed of one $2$-simplex ($\left\{a,b,c\right\}$), three $1$-simplices ($\{a,b\}, \{b,c\}$ and~$\{a,c\}$) and three $0$-simplices ($\{a\}, \{b\}$ and~$\{c\}$).

\textbf{Important notations.} In this work, the symbol~\C denotes a non-empty $n$-complex, with $n\in\mathbb{N}$. The set of all subsets of $\C$ is denoted by \PC. Any subset of~$\C$ that is also a complex is called a \emph{subcomplex (of~$\C$)}. We denote by~$\subcomplexes$ the set of all subcomplexes of~$\C$.

If~$X$ is a subset of~$\C$, we denote by~$\overline{X}$ the \emph{complement} of~$X$ (in \C):~$\overline{X}=\C \backslash X$. The complement of a subcomplex of~$\C$ is usually not a subcomplex. Any subset $X$ of~$\C$ whose complement~$\overline{X}$ is a subcomplex is called a \emph{star}. We denote by~$\stars$ the set of all stars in $\C$. The intersection~$\subcomplexes\,\cap\,\stars$ is non-empty since it always contains at least~$\emptyset$ and~$\C$.

In the domain of simplicial complexes, some operators are well known, such as the \emph{closure} and \emph{star}~\citep{Jaenich1984}. We define the closure $\hat{x}$ and the star $\check{x}$ of a simplex $x$ as:

\begin{align}
  \forall x \in \C,\,\hat{x}=&\left\{ y \st y \subseteq x, y \neq \emptyset \right\}\label{eq:hatx}\\
  \forall x \in \C,\,\check{x}=&\left\{ y\in \C \st x \subseteq y\right\}\label{eq:checkx}
\end{align}

In other words, the closure operator gives as result the set of all simplices that are subsets of the simplex $x$, and the star gives as result the set of all simplices of $\C$ that contain the simplex $x$. These operators can be easily extended to sets of simplices. The operators~$Cl:\PC \rightarrow \PC $ and~$St: \PC \rightarrow \PC$ are defined by:

\begin{align}
  \forall X \in \PC, \ Cl =&\bigcup \{\hat{x} \st x\in X\}\label{eq:ClSetPC}\\
  \forall X \in \PC ,\ St =&\bigcup\{ \check{x} \st x\in X\} \label{eq:StSetPC}
\end{align}

\subsection{Mathematical morphology}
\label{sec:mm}
We approach mathematical morphology through the framework of lattices~\citep{Ronse1990}. We start with the concept of partially ordered set. It is composed by a set and a binary relation. The binary relation is defined only between certain pairs of elements of the set, representing precedence, and must be reflexive, antisymmetric and transitive. 

A \emph{lattice} is a partially ordered set with a least upper bound, called \emph{supremum}, and a greatest lower bound, called
\emph{infimum}. For instance, the set $\mathcal{P}(S)=\{\{a,b,c\}, \{a,b\}, \{a,c\}, \{b,c\}, \{a\}, \{b\}, \{c\},\emptyset\}$, that is the power set of the set $S=\{a,b,c\}$, ordered by the inclusion relation, is a lattice. The supremum of two elements of this lattice is given by the union operator and the infimum by the intersection operator. This lattice can be denoted by $\langle\mathcal{P}(S),\bigcup,\bigcap,\subseteq\rangle$.   

In mathematical morphology (see, \eg, \cite{Ronse2010}), any operator that associates elements of a lattice $\L[1]$ to elements of a lattice $\L[2]$ is called a \emph{dilation} if it commutes with the supremum. Similarly, an operator that commutes with the infimum is called an \emph{erosion}.

Let~$\L[1]$ and~$\L[2]$ be two lattices whose order relations and suprema are denoted by~$\leq_1$,~$\leq_2$, $\vee_1$, and~$\vee_2$. Two operators $\alpha:\DomImL{2}{1}$ and $\beta:\DomImL{1}{2}$ form an \emph{adjunction} $(\beta,\alpha)$ if $\alpha(a)\leq_{1} b \ \iff\ a \leq_{2} \beta(b)$ for every element $a$ in $\mathcal{L}_{2}$ and $b$ in $\mathcal{L}_{1}$. It is well known (see, \eg,~\cite{Ronse2010}) that, given two operators~$\alpha$ and~$\beta$, if the pair~$(\beta,\alpha)$ is an adjunction, then~$\beta$ is an erosion and~$\alpha$ is a dilation. Furthermore, if~$\alpha$ is a dilation, there is a unique erosion~$\beta$, called the {\em adjoint of~$\alpha$}, such that~$(\beta,\alpha)$ is an adjunction. This erosion is characterized by:

\begin{equation}
  \forall a \in \L[1],\, \beta(a)=\vee_2 \left\{ b \in \L[2] \st \alpha(b) \leq_1 a \right\}
  \label{eq:AdjDef}
\end{equation}

In certain cases, we will denote the adjoint operator of an operator $\alpha$ by $\alpha\adjoint$, to explicit the relationship between them.

In mathematical morphology, an operator $\alpha$, acting from a lattice $\L_1$ to $L_2$, that is increasing ($\forall a,b\in\L_1,\ a\leq b \implies \alpha(a)\leq \alpha(b)$) and idempotent ($\forall a\in\L_1,\ \alpha(a)=\alpha(\alpha(a))$) is a \emph{filter}. If a filter is anti-extensive ($\forall a\in\L_1,\ \alpha(a)\leq a$) it is called an \emph{opening}. Similarly, an extensive filter ($\forall a\in\L_1,\ a\leq \alpha(a)$) is called a \emph{closing}. 

One way of obtaining openings and closings is by combining dilations and erosions~\citep{Ronse2010}. Let $\alpha:\DomIm{\L}{\L}$ be a dilation. Then the operator $\zeta=\alpha\adjoint\alpha$ is a closing and the operator $\psi=\alpha\alpha\adjoint$ is an opening. Both operators act on $\L$.

A family of openings $\Psi=\{\psi_{\lambda} \st \lambda \in \mathbb{N}\}$ acting on $\L$, is a \emph{granulometry} if, given two
positive integers $i$ and $j$, we have $i \geq j \implies \psi_{i}(a) \subseteq \psi_{j}(a)$, for any $a\in\L$~\citep{Ronse2010}. Similarly, a family of closings $Z=\{\zeta_{\lambda} \st \lambda \geq 0\}$, is a \emph{anti-granulometry} if, given two positive integers $i$ and $j$, we have $i \leq j \implies \zeta_{i}(a) \subseteq \zeta_{j}(a)$, for any $a\in\L$.

A family of filters $\{ \alpha_\lambda , \lambda \in \N\}$ is a family of \emph{alternating sequential filters} if, given two positive integers $i$ and $j$, we have $i > j \implies \alpha_{i}  \alpha_{j} = \alpha_{i}$. 

Let $\Psi=\{\psi_{\lambda},\,\lambda\in\N\}$ be a granulometry and $Z=\{\zeta_{\lambda},\,\lambda\in\N\}$ be an anti-granulometry. We can construct two alternating sequential filters by composing operators from both families. Let $i\in\N$ and $a\in\L$:

\begin{align}
	\nu_i(a)= &\paren{\psi_{i}\zeta_{i}}\paren{\psi_{i-1}\zeta_{i-1}}\dots\paren{\psi_{1}\zeta_{1}} (a) \label{eq:asfDef}\\
	\nu_i'(a)= &\paren{\zeta_{i}\psi_{i}}\paren{\zeta_{i-1}\psi_{i-1}}\dots\paren{\zeta_{1}\psi_{1}} (a) \label{eq:asfAltDef}
\end{align}

\section{Dimensional operators}
\label{sec:DimOps}
In~\cite{Dias2011}, we introduced four new basic operators that act on simplices of given dimensions. These operators can be composed into new operators which behavior can be finely controlled. We proceed with a brief reminder of their definition and explore some new properties. 

We start by introducing a new notation that allows only simplices of a given dimension to be retrieved. Let~$X \subseteq \C$ and let~$i \in [0,n]$, we denote by~$\dimension{X}{i}$ the set of all $i$-simplices of~$X$:~$\dimension{X}{i} = \{x \in X \st dim(x) = i\}$. In particular, $\C[i]$ is the set of all $i$-simplices of $\C$. We denote by $\PC[i]$ the set of all subsets of~$\C[i]$. 

Let $i\in\N$ such that $i\in [0,n]$. The structure $\left\langle \PC[i], \bigcup,\bigcap,\subseteq \right\rangle$ is a lattice. 

  \begin{defn}
  \label{def:DimOp}
  Let~$i,j \in \mathbb{N}$ such that~$0\leq i < j \leq n$, $X\in\PC[i]$ and $Y\in\PC[j]$. We define the operators~$\dij$ and~$\eij$ acting from~$\PC[i]$ into~$\PC[j]$ and the operators~$\dji$ and~$\eji$ acting from~$\PC[j]$ into~$\PC[i]$ by:

  \begin{align}
		\dij(X) &= \{x \in \C[j] \st\exists y \in X, y \subseteq x \}\\
		\eij(X) &= \{x \in \C[j] \st\forall y \in \C[i],y\subseteq x \Rightarrow y \in X\}\\
		\dji(Y) &= \{x \in \C[i] \st\exists y \in Y, x \subseteq y \}\\
		\eji(Y) &= \{x \in \C[i] \st\forall y \in \C[j], x \subseteq y \Rightarrow y \in Y\}\!
  \end{align}
\end{defn}

In other words,~$\dij(X)$ is the set of all $j$-simplices of~$\C$ that include an $i$-simplex of~$X$,~$\dji(X)$ is the set of all~$i$-simplices of~$\C$ that are included in a $j$-simplex of~$X$,~$\eij(X)$ is the set of all~$j$-simplices of~$\C$ whose subsets of dimension~$i$ all belong to~$X$, and~$\eji(X)$ is the set of all~$i$-simplices of~$\C$ that are not contained in any~$j$-simplex of~$\overline{X}$. 

The dimensional operators can also be recovered using the classical star and closure operators:

\begin{proper}
    \label{p:DimOpAlt}
    We have:
    \begin{enumerate}
      \item{$\forall X\subseteq\C[i],\,\dij(X)=\Dim{\left[St(X)\right]}{j}$;}
      \item{$\forall X\subseteq\C[j],\,\dji(X)=\Dim{\left[Cl(X)\right]}{i}$;}
      \item{$\forall X\subseteq\C[i],\,\eij(X)=\Dim{\left[\ov{St\paren{\ov{X}}}\right]}{j}$;}
      \item{$\forall X\subseteq\C[j],\,\eji(X)=\Dim{\left[\ov{Cl\paren{\ov{X}}}\right]}{i}$.}
    \end{enumerate}
\end{proper}

The dimensional operators can be useful when the considered data is associated only with simplices of a given dimension of the complex, which is fairly common. In this situation, these operators can be used to propagate the values to the other dimensions of the complex, or even filter the values directly, depending on the application. However, since the objective of this work is to find interesting operators acting on subcomplexes, we mostly use these operators as building blocks to define new operators. The following adjunction property can be proved by constructing the adjoint erosion of the dilation operators and verifying that they correspond to the provided erosion, the properties regarding duality are trivial results from property~\ref{p:DimOpAlt}.

\begin{proper}
  \label{p:DimOps}
  Let~$i,j \in \mathbb{N}$ such that~$0\leq i < j \leq n$.
  \begin{enumerate}
   \item{The pairs~$(\eij, \dji)$ and~$(\eji, \dij)$ are adjunctions; }
   \item{The operators~$\dij$ and~$\eij$ are dual of each  other:\\ $\forall X \subseteq \C[i],\ \eij (X) =  \C[j] \setminus \dij(\C[i] \setminus X)$;}
   \item{The operators~$\dji$ and~$\eji$ are dual of each other:\\ $\forall X \subseteq \C[j],\ \eji(X) =\C[i] \setminus \dji(\C[j] \setminus X)$.  }
  \end{enumerate}
\end{proper}

We can use the dimensional operators from definition~\ref{def:DimOp} to define new operators, leading to new dilations, erosions, openings, closings and alternating sequential filters. Before we start composing these operators, let us consider the following results, that can guide the exploration of new compositions.

\begin{proper}
  \label{p:CompOp}
    Let $i,j,k\in\N$ such that $0\leq i < j < k \leq n$.
    \begin{enumerate}
    \item{$\forall X\subseteq\PC[i],\,\dilp{j}{k}\dilp{i}{j}(X)=\dilp{i}{k}(X)$;}
    \item{$\forall X\subseteq\PC[i],\,\erop{j}{k}\erop{i}{j}(X)=\erop{i}{k}(X)$;}
    \item{$\forall X\subseteq\PC[k],\,\dilm{j}{i}\dilm{k}{j}(X)=\dilm{k}{i}(X)$;}
    \item{$\forall X\subseteq\PC[k],\,\erom{j}{i}\erom{k}{j}(X)=\erom{k}{i}(X)$.}
    \end{enumerate}
\end{proper}

Property~\ref{p:CompOp} states that any composition of the same operator is equivalent to the operator acting from the initial to the final dimension. The proof of this property can be done by contradiction, where if $\dilp{j}{k}\dilp{i}{j}(X)\not=\dilp{i}{k}(X)$ is true, our space is not a simplicial complex.

To explore the possible combinations of the operators from definition~\ref{def:DimOp}, we start by considering only operators acting on the same dimension. The following property can be deduced from property~\ref{p:DimOpAlt}:

\begin{proper}
  \label{p:OpInd}	
    Let $i,j,k\in\N$ such that $0\leq i < j < k \leq n$.
    \begin{enumerate}
    \item{$\forall X\subseteq\PC[i],\,\dilm{j}{i}\dilp{i}{j}(X)=\dilm{k}{i}\dilp{i}{k}(X)$;}
    \item{$\forall X\subseteq\PC[i],\,\erom{j}{i}\erop{i}{j}(X)=\erom{k}{i}\erop{i}{k}(X)$;}
    \item{$\forall X\subseteq\PC[i],\,\erom{j}{i}\dilp{i}{j}(X)=\erom{k}{i}\dilp{i}{k}(X)$;}
    \item{$\forall X\subseteq\PC[i],\,\dilm{j}{i}\erop{i}{j}(X)=\dilm{k}{i}\erop{i}{k}(X)$.}
    \end{enumerate}
\end{proper}

Property~\ref{p:OpInd} states that the result of the compositions of dilations and erosions that use a higher intermediary dimension is independent of the dimension chosen. Therefore, we can obtain only one dilation, one erosion, one opening and one closing using those compositions. However, this is not entirely true when we consider a lower dimension as intermediary dimension for the compositions. The following property can be deduced from property~\ref{p:DimOpAlt}:

\begin{proper}
  \label{p:FamilyDownUp}
    Let $i,j,k\in\N$ such that $0\leq i < j < k \leq n$.
    \begin{enumerate}
    \item{$\forall X\in\PC[k],\,\dilp{i}{k}\dilm{k}{i}(X)\supseteq\dilp{j}{k}\dilm{k}{j}(X)$;}
    \item{$\forall X\in\PC[k],\,\erop{i}{k}\erom{k}{i}(X)\subseteq\erop{j}{k}\erom{k}{j}(X)$;}
    \item{$\forall X\in\PC[k],\,\erop{i}{k}\dilm{k}{i}(X)= \erop{j}{k}\dilm{k}{j}(X)$;}
    \item{$\forall X\in\PC[k],\,\dilp{i}{k}\erom{k}{i}(X)= \dilp{j}{k}\erom{k}{j}(X)$.}
    \end{enumerate}
\end{proper}

Property~\ref{p:FamilyDownUp} states that compositions from dilations and erosions using a lower intermediary dimension are equal, independent of the chosen dimension and that compositions of only dilations and erosions are related, but not always equivalent.

\subsection{Extension to weighted complexes}

In this section, we extend the operators we defined to weighted simplicial complexes.  Let $k_{min}$ and $k_{max}$ be two distinct, positive integers. We define the set $\K$ as the set of the integers between these two numbers, $\K=\{k\in\N \st k_{min}\leq k\leq k_{max}\}$. Now, let $M$ be a map from $\C$ to $\K$, that associates an element of $\K$ to every element of the simplicial complex $\C$. Let $x\in\C$, in this work, $M(x)$ is called the \emph{value} of the simplex $x$.

We can extend the notion of subcomplexes and stars to the domain of weighted complexes. A map~$M$ from~$\C$ in~$\K$ is a {\em simplicial stack} (see \cite{Cousty2009c}) if the value of each simplex is smaller than or equal to the value of the simplices it includes, $\forall x\in X,\,\forall y\subseteq x, M(x)\geq M(y)$. On the other hand, when the comparison is reversed, when $\forall x\in \C,\,\forall y\subseteq x, M(x)\leq M(y)$, we say that~$M$ is a {\em starred stack}. The {\em dual $\overline{M}$} of a map $M$ is defined using the value $k_{max}$: $\forall x\in \C, \overline{M}(x)=k_{max}-M(x)$.

Let~$M$ be a map from an arbitrary set~$E$ in~$\K$ and let $k\in\K$. We denote by $M[k]$ the set of elements in~$E$ with value greater than or equal to $k$, $M[k]=\{x\in E \st M(x)\geq k\}$. This set is called the \emph{$k$-threshold} of $M$.

The following lemma, which can be easily proved from the definitions, clarifies the links between stars, complexes, and the $k$-thresholds of simplicial stacks and starred stacks. 
\begin{lemma}
The following relations hold true:
  \begin{enumerate}
    \item $M$  is a simplicial stack $\iff$ $\forall k \in \K,\,M[k]\in\subcomplexes$;
    \item $M$  is a starred stack $\iff \forall k \in \K,\,M[k]\in\stars$;
    \item $M$  is a simplicial stack  $\iff \ov{M}$ is a starred stack.
  \end{enumerate}
\end{lemma}

We approach the problem of extending the dimensional operators to weighted complexes using threshold decomposition and stack reconstruction (see, \eg~\cite{Serra1982}). The main idea of this method is that, if the considered operator is increasing, we can apply it to each $k$-threshold and then combine the results to obtain the final values. More precisely, let $E_1$ and $E_2$ be two sets and $\alpha$ an increasing operator from~$\mathcal{P}(E_1)$ to~$\mathcal{P}(E_2)$, the {\em extended stack operator of~$\alpha$}, also denoted by~$\alpha$, is:

\begin{align}
  \forall &M:E_1 \rightarrow \K,  \forall x \in E_2,\nonumber\\
	&[\alpha(M)](x) =\max\{ k \in \K \st x \in \alpha(M[k])  \}
  \label{eq:stack}
\end{align}

As erosions and dilations, the dimensional operators are increasing. Thus, they can be extended to maps. Their extended stack operators are characterized by the following property.

\begin{proper}
  Let~$i,j \in \K$ such that~$i \leq j$, let $M_i: \PC[i] \rightarrow \K $ and~$M_j: \PC[j] \rightarrow \K$. 
  \label{p:DimOpGray}
  \begin{enumerate}
    \item $\begin{dcases}\forall x \in \C_i,\\ [\dji(M_j)](x)=\max_{y \in \C[j]}  \{M_j(y) \st x \subseteq y\};\end{dcases}$
    \item $\begin{dcases}\forall x \in \C_i,\\ [\eji(M_j)](x)=\min_{y \in \C[j]} \{M_j(y) \st x \subseteq y\});\end{dcases}$
    \item $\begin{dcases}\forall x \in \C_j,\\ [\dij(M_i)](x)=\max_{y \in \C[i]} \{M_i(y) \st  y \subseteq x\});\end{dcases}$
    \item $\begin{dcases}\forall x \in \C_j,\\ [\eij(M_i)](x)=\min_{y \in \C[i]} \{M_i(y) \st y \subseteq x\}).\end{dcases}$
  \end{enumerate}
\end{proper} 
 
\subsection{Revisiting the related work}
\label{sec:rev}

In section~\ref{sec:DimOps} we defined operators acting between specific dimensions of the complex. Here, we use these operators, considering in particular property~\ref{p:DimOpGray}, to express operators from the literature.

We start by the classical star and closure operators. Let $X\subseteq\C$.

\begin{align}
  St(X)= &\bigcup\left\{\dij(\dimension{X}{i}) \st i,j\in\N ,i\leq j\right\}\\
  Cl(X)= &\bigcup\left\{\dji(\dimension{X}{j}) \st i,j\in\N ,i\leq j\right\}
\end{align}

\cite{Vincent1989} defined operators acting on a vertex weighted graph $(V,E, f)$, where $V$ is a finite set (of vertices), $E$ is a set of unordered pairs of~$V$, called edges, and~$f$ is a map from~$V$ in~$\K$. By abuse of terminology,~$f$ is called a weighted graph.

Let $v \in V$, the set of neighbors of a vertex $v$ is given by $N_E(v)=\left\{v'\in V \st \{v,v'\} \in E\right\}$. The dilated graph $\Gamma(f)$ and the eroded graph $\Gamma^0(f)$ of the graph~$f$ are given, for any vertex~$v$, by:
\begin{enumerate}
  \item $[\Gamma(f)](v) = \max\left\{f(v') \st v'\in N_E(v)\cup\left\{v\right\}\right\}$;
  \item $[\Gamma^0(f)](v) = \min\left\{f(v') \st v'\in N_E(v) \cup \left\{v\right\}\right\}$.
\end{enumerate}

In other words, these operators replace the value of each vertex with the maximum (or minimum) value of its neighbors, as morphological operators often do. To be able to draw a parallel between these operators and the dimensional operators presented in this work, let us consider the 1-complex~$\C$ defined as the union of the vertex and edge sets of the graph~$G$ : $\C = V \cup E$. Observe then that~$C_0 = V$ and~$\C_1 = E$ and that~$f$ is a map weighting the 0-simplices of~$\C$. Using the dimensional operators, we can recover the operators from \cite{Vincent1989}:

\begin{align}
  \Gamma(f)   =  & \dilm{1}{0}\dilp{0}{1}(f)\\
  \Gamma^0(f) =  & \erom{1}{0}\erop{0}{1}(f)
\end{align}

From these basic dilations and erosions \cite{Vincent1989} derives several interesting operators, which, thanks to the previous relations, can be recovered using the operators of this article. 

So far the graphs were used only to provide structural information about the considered space. By considering the edges and vertices in an uniform way, allowing the propagation of the values also to the edges of the graph, both Cousty~\etal~\citep{Cousty2009} and Meyer and Stawiaski~\citep{Meyer2009} obtained new operators.

Cousty~\etal~\citep{Cousty2009,Cousty2013} considered a graph $\G=(\G^{\bullet},\G^{\times})$. For any $X^{\times}\subseteq\G^{\times}$ and $Y^{\bullet}\subseteq\G^{\bullet}$, the operators $\varepsilon^{\times}$, $\delta^{\times}$, $\varepsilon^{\bullet}$ and $\delta^{\bullet}$ are defined by:

\begin{align}
  \varepsilon^{\times}(Y^{\bullet})=&\left\{e_{x,y}\in\G^{\times}\st x\in Y^{\bullet} \text{ and } y\in Y^{\bullet}\right\}\\
  \delta^{\times}(Y^{\bullet})= &\left\{e_{x,y}\in\G^{\times} \st x\in Y^{\bullet} \text{ or } y \in Y^{\bullet}\right\}\\ 
  \varepsilon^{\bullet}(X^{\times})=&\left\{x\in\G^{\bullet}\st \forall e_{x,y} \in\G^{\times}, e_{x,y}\in X^{\times}\right\}\\
  \delta^{\bullet}(X^{\times})=&\left\{x\in\G^{\bullet}\st \exists e_{x,y}\in X^{\times}\right\}
\end{align}

If the considered space $\C$ is the 1-complex~$\C= \G^{\bullet} \cup \G^{\times}$, using the dimensional operators, we have:

\begin{align}
  \varepsilon^{\times}(Y^{\bullet})  = &\erop{0}{1}(Y^{\bullet})\\
  \delta^{\times}(Y^{\bullet})       = &\dilp{0}{1}(Y^{\bullet})\\
  \varepsilon^{\bullet}(X^{\times}) = &\erom{1}{0}(X^{\times})\\
  \delta^{\bullet}(X^{\times})      = &\dilm{1}{0}(X^{\times}) 
\end{align}

Later, in~\cite{Cousty2013}, Cousty~\etal extended these operators to weighted graphs, but the relations presented here are still true. Meyer, Angulo and Stawiaski~\citep{Meyer2007,Meyer2009} also defined operators capable of dealing with weighted graphs. They consider the space as a graph $G=(N,E)$, where $N=\left\{n_1, n_2, \dots, n_{|N|}\right\}$ is the set of vertices and $E=\left\{e_{ij} \st i,j \in \mathbb{N}^{+}, 0<i<j\leq|N|\right\}$ is the set of edges. For two functions~$n$ and~$e$ weighting the vertices and edges of~$G$, they consider the following operators:

\begin{align}
  \left[\varepsilon_{en}n\right]_{ij}  =& n_i \wedge n_j\\
  \left[\delta_{ne}e     \right]_{i\ } =& \bigvee\nolimits_{k \text{ neighbors of } i} \left\{e_{ik}\right\}\\
  \left[\varepsilon_{ne}e\right]_{i\ } =& \bigwedge\nolimits_{k \text{ neighbors of } i} \left\{e_{ik}\right\}\\
  \left[\delta_{en}n     \right]_{ij}  =& n_i \vee n_j
\end{align}
If the considered space $\C$ is the 1-complex $\C= \{1, \ldots, |N|\} \cup \{\{i,j\} \st e_{ij} \in E\}$, using the dimensional operators
from definition~\ref{def:DimOp}, we have:

\begin{align}
  \left[\varepsilon_{en}n\right]_{ij}  =&{[\erop{0}{1}(n)](\{i,j\})}\\
  \left[\delta_{en}n\right]_{ij} =&{[\dilp{0}{1}(n)](\{i,j\})}\\
  \left[\varepsilon_{ne}e\right]_{i}  =&{[\erom{1}{0}(e)](\{i\})}\\
  \left[\delta_{ne}e\right]_{i}  = &{[\dilm{1}{0}(e)](\{i\})}.
\end{align}

Meyer, Angulo, and Stawiaski~\citep{Meyer2007,Meyer2009} defined several operators based on the four presented above, all of them are recoverable by the dimensional operators. They also defined operators that rely on the particular structure of the hexagonal grid and cannot be easily expressed using our operators.

\subsection{Morphological operators on $\subcomplexes$ using a higher intermediary dimension}

In this section, we define new operators, acting on subcomplexes, whose result is a complex of the same dimension of its argument, using an higher intermediary dimension, exploring the effects of property~\ref{p:OpInd}. For instance, if we consider a complex $X$ of dimension $i$, with $i\in\N,\,0<i\leq n$, we would like the dilation of $X$ to be also an $i$-complex. To that end, the operators proposed next act independently on each dimension of the complex:

\begin{defn}
  We define:
  \label{def:dilG}
  \label{def:eroG}  
  \begin{align}
    \forall X&\in\subcomplexes,\,\dilG (X)=\nonumber\\
		& \left\{ \bigcup\nolimits_{i\in[0\dots(n-1)]} \dilm{i+1}{i}\dilp{i}{i+1}(\dimension{X}{i}) \right\} \bigcup\nonumber\\
		&\left\{\dilp{n-1}{n}  \dilm{n}{n-1}(\dimension{X}{n}) \right\} \label{eq:dilG}\\
    \forall X&\in\subcomplexes,\,\eroG(X) =\nonumber\\
		& Cl\adjoint \left(\left\{ \bigcup\nolimits_{i\in[0\dots(n-1)]} \erom{i+1}{i}\erop{i}{i+1}(\dimension{X}{i}) \right\}\right. \bigcup  \nonumber\\
    &\left\{\erop{n-1}{n}  \erom{n}{n-1}(\dimension{X}{n}) \right\}\bigg)  \label{eq:eroG}
  \end{align}
\end{defn}

As expected, the set $\Dim{(\dilG(X))}{i}$, made of the $i$-simplices of $\dilG(X)$, depends only on the set $\Dim{X}{i}$, made of the $i$-simplices of $X$. Intuitively, for $i < n$, the set $\Dim{(\dilG(X))}{i}$ contains all $i$-simplices of $\C$ that either belong to $\Dim{X}{i}$ or are contained in a $(i+1)$-simplex that includes an $i$-simplex of $\Dim{X}{i}$. For $i=n$, the operator will return all $n$-simplices that contains an $(n-1)$-simplex of X. 

\begin{figure}
  \centering 
  \subfigure[$Y$] {\includegraphics[width=0.2\textwidth]{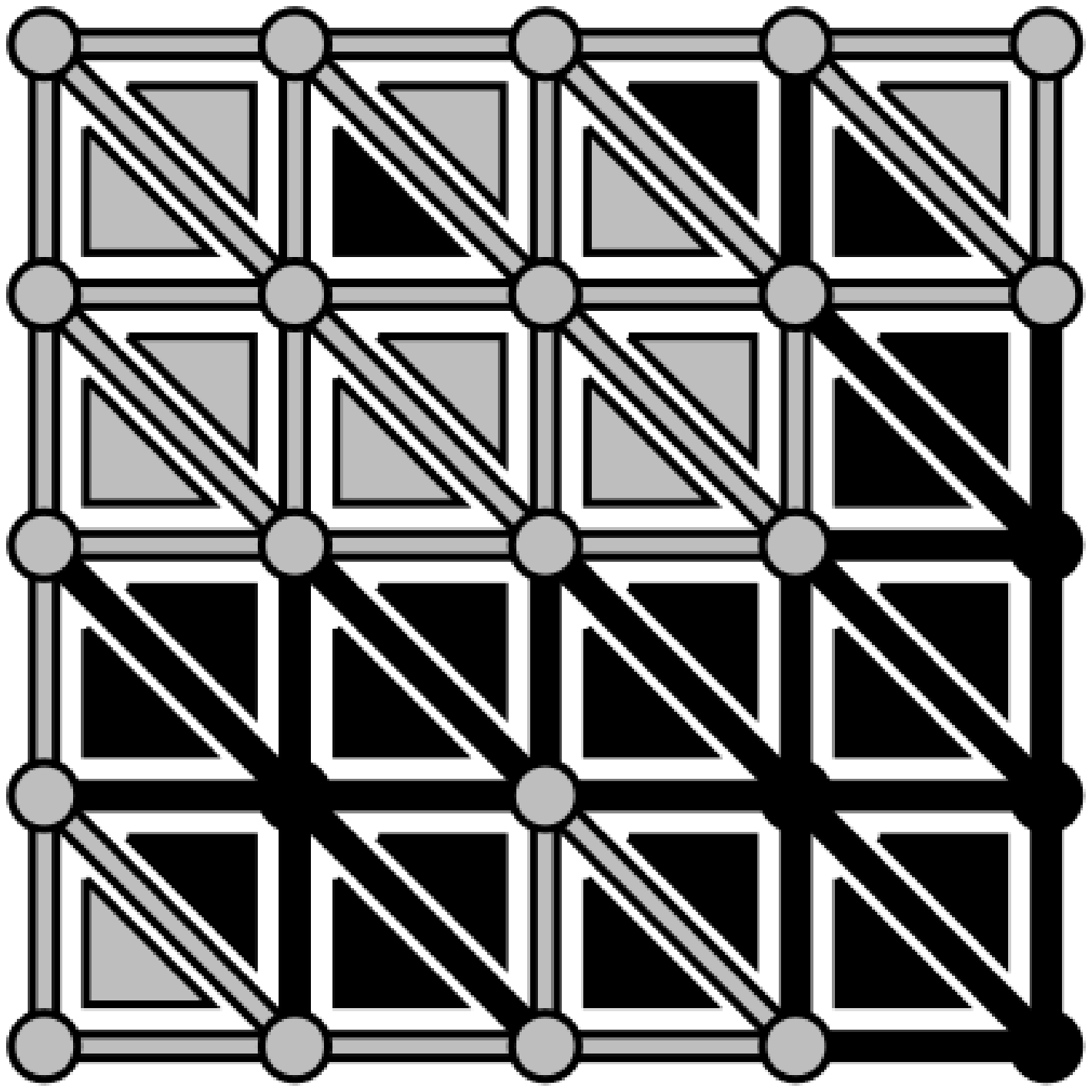}}
  \\
  \subfigure[$\dilation(Y)$]{\includegraphics[width=0.2\textwidth]{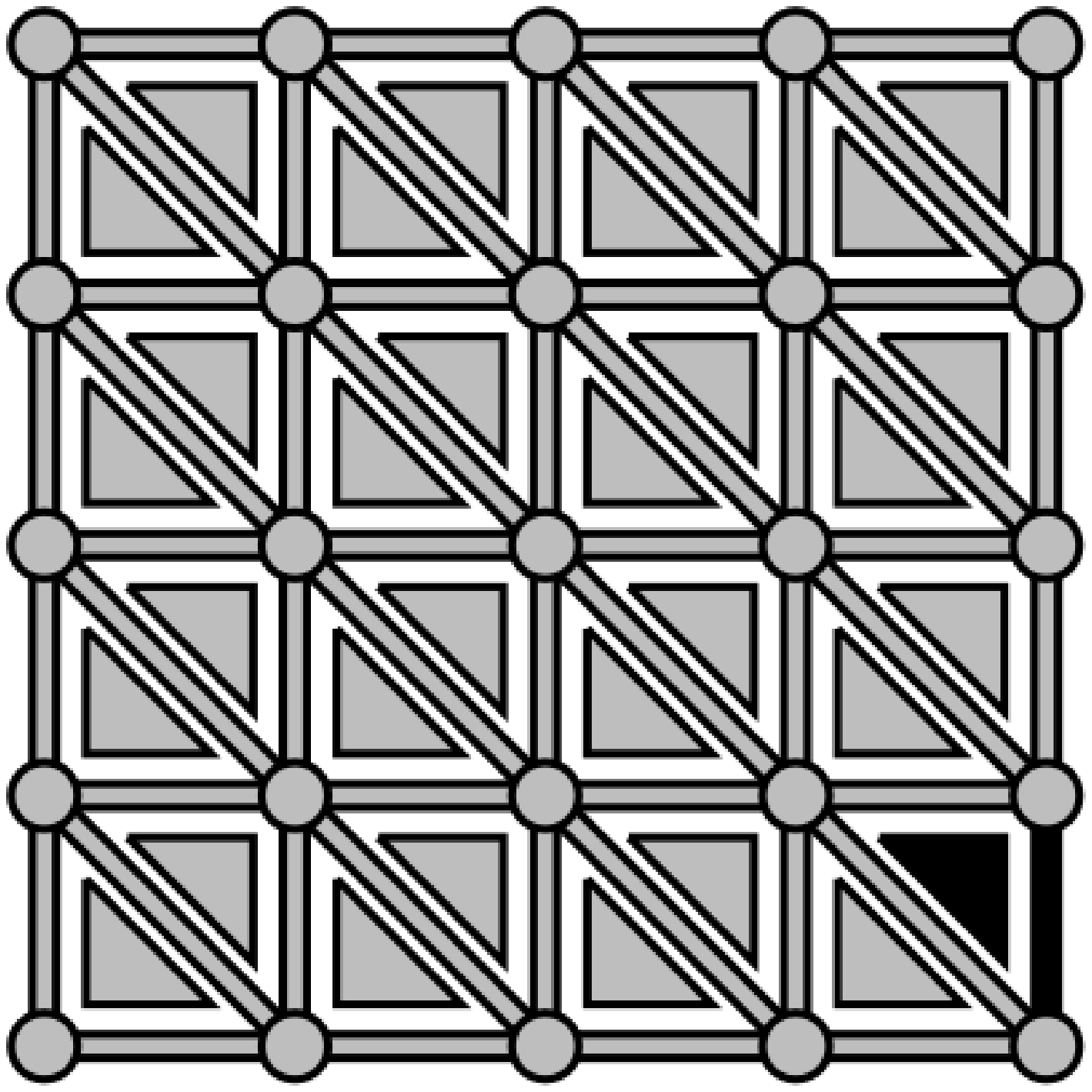}}
  \subfigure[$\dilG(Y)$]{\includegraphics[width=0.2\textwidth]{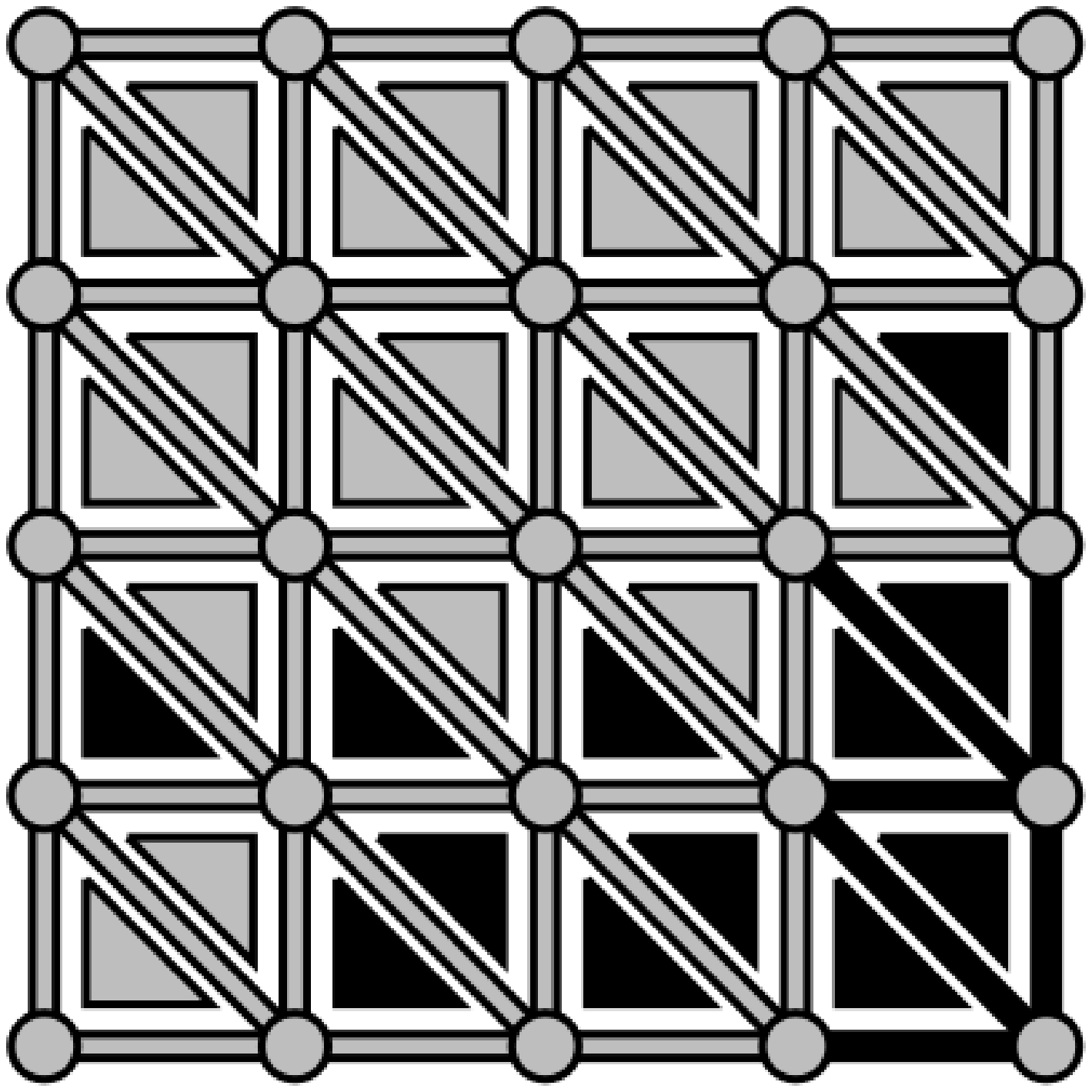}}
  \subfigure[$\erosion(Y)$]{\includegraphics[width=0.2\textwidth]{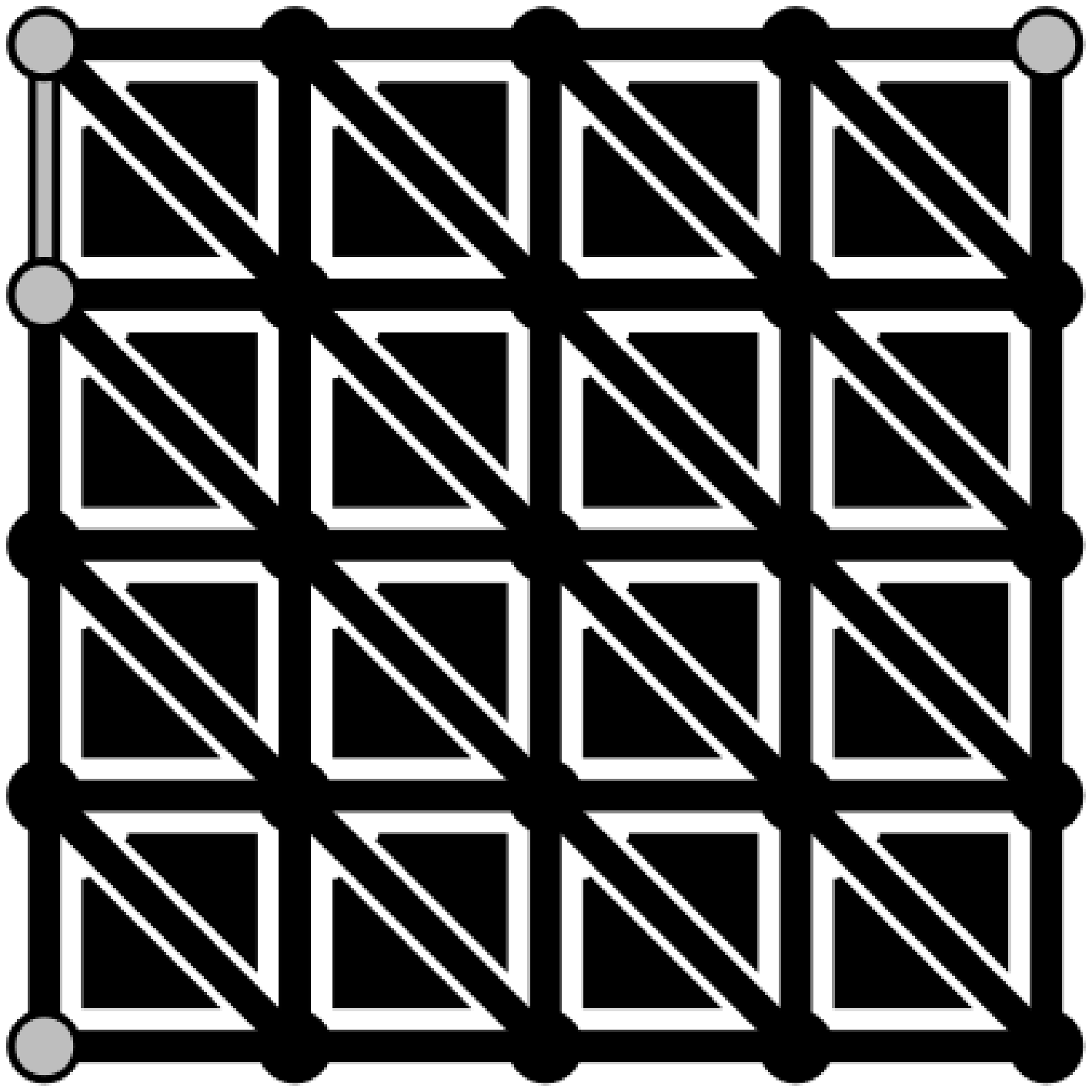}}
  \subfigure[$\eroG(Y)$]{\includegraphics[width=0.2\textwidth]{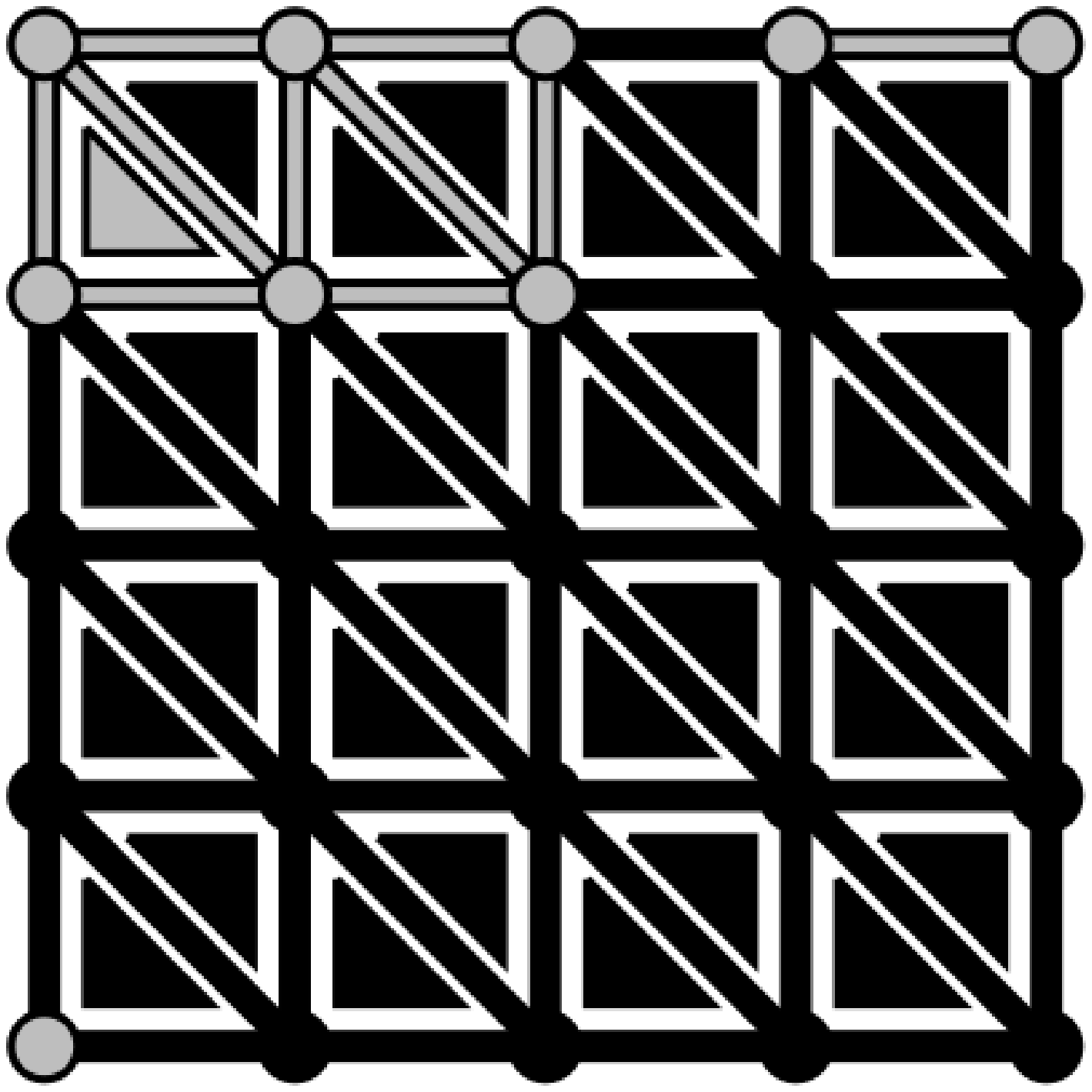}}
	\caption{Illustration of the operators $\dilG$ and $\eroG$ [see text].\label{fig:EroDilG}}
\end{figure}

Some results of the operators $\dilG$ and $\eroG$, along with the results of the operators $\dilation$ and $\erosion$ introduced by~\cite{Dias2011}, are depicted as gray simplices in the figure~\ref{fig:EroDilG}. As expected, these operators result in a subcomplex more similar to the argument. The dilation included less simplices into the set, while the erosion removed less simplices of the set.

It can easily be proven that the operators $\eroG$ and $\dilG$ act on $\subcomplexes$ and form an adjunction. Therefore, we can compose them to define new operators. 

\begin{defn}
  \label{def:OpClG}
  Let $i\in\N$. We define:
  \begin{align}
    \opG[i]=&\paren{\dilG}^{i}  \paren{\eroG}^{i}  \\
    \clG[i]=&\paren{\eroG}^{i}  \paren{\dilG}^{i}  
  \end{align}
\end{defn}

\begin{figure}
  \centering
  \subfigure[$Y$]{\includegraphics[width=0.2\textwidth]{Y.eps}}
  \subfigure[$Z$]{\includegraphics[width=0.2\textwidth]{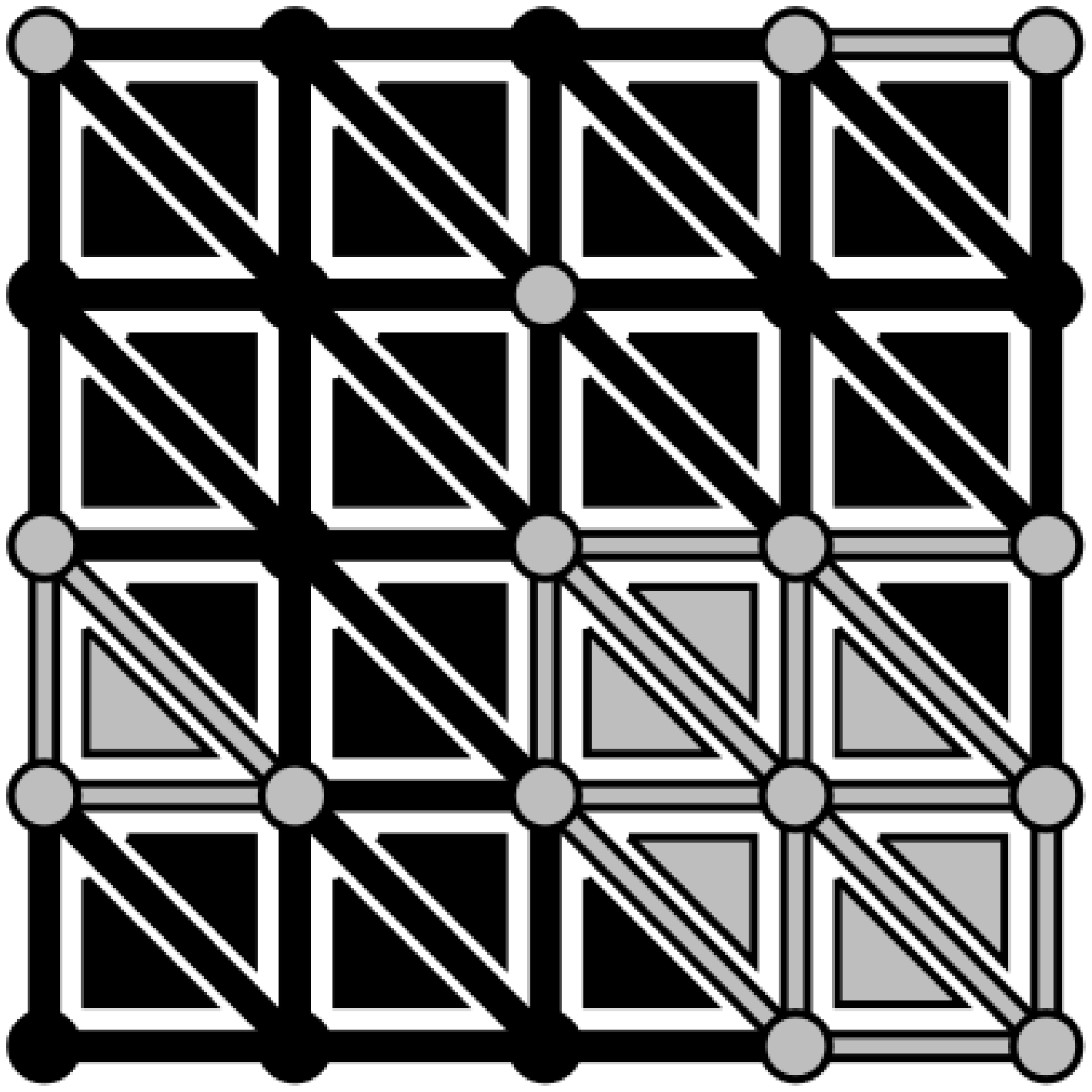}}
  \subfigure[$\phi^{\uptodownarrow}_{1}(Y)$]{\includegraphics[width=0.2\textwidth]{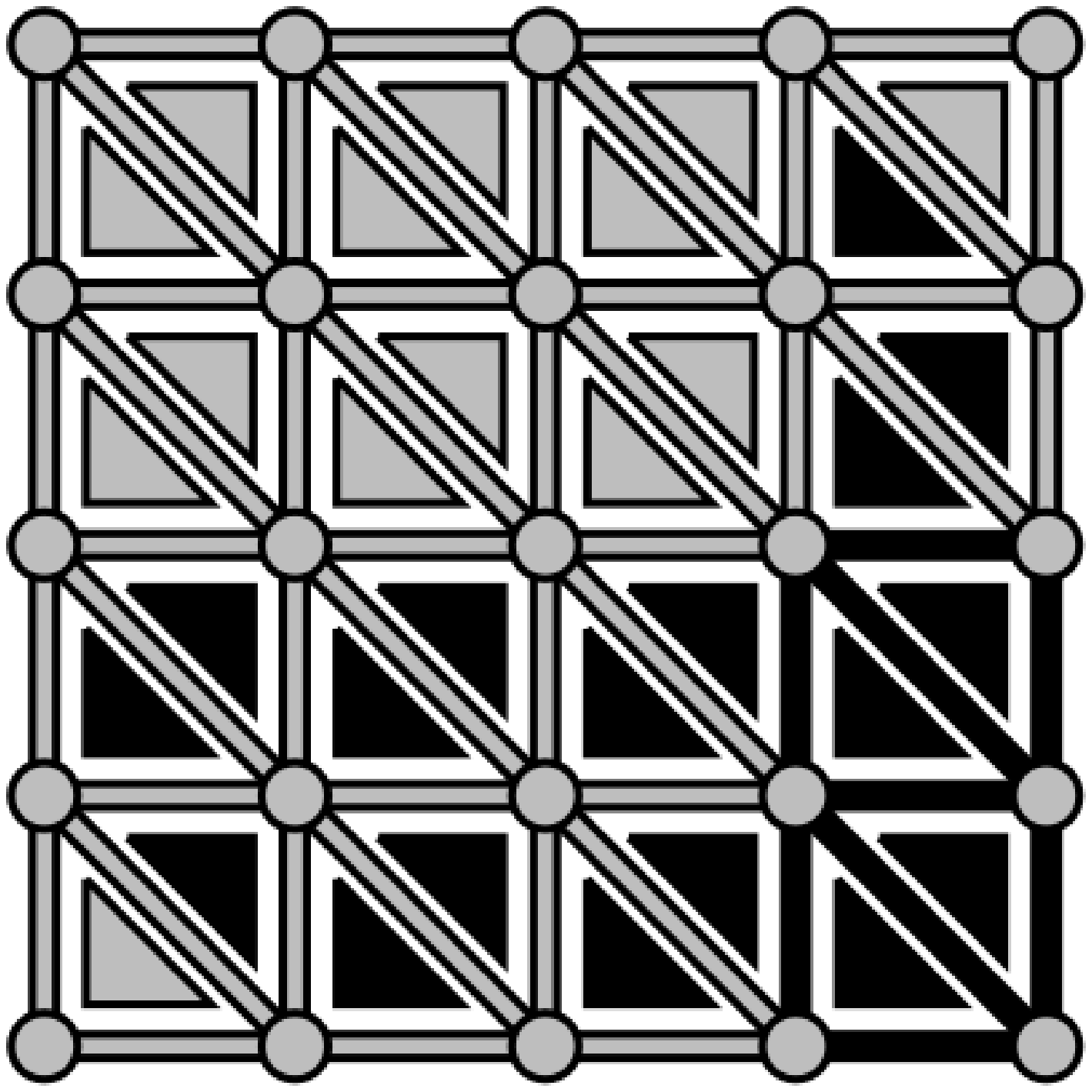}}
  \subfigure[$\gamma^{\uptodownarrow}_{1}(Z)$]{\includegraphics[width=0.2\textwidth]{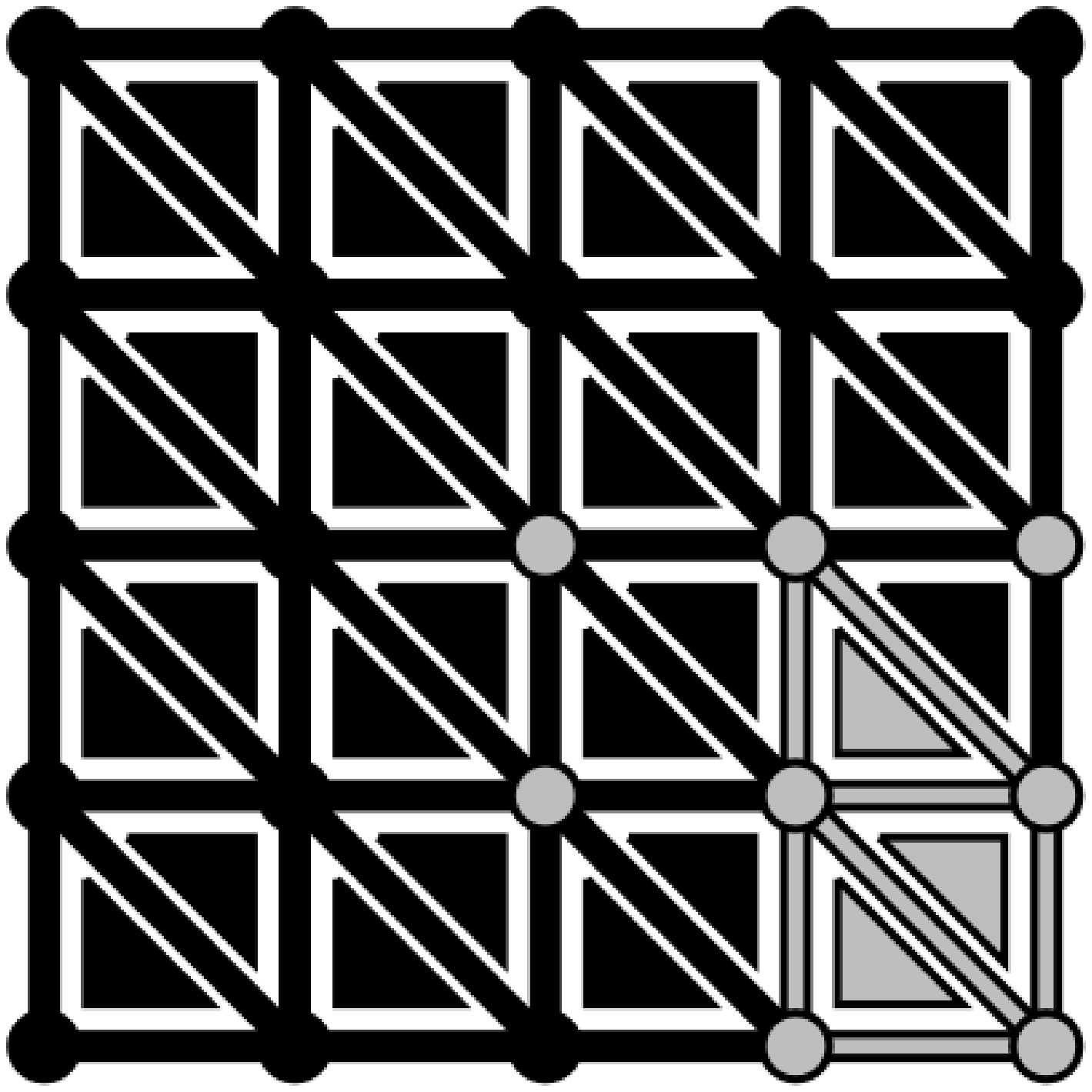}}
  \caption{Illustration of operators $\opG[i]$ and $\clG[i]$.\label{fig:OpClG}}
\end{figure}

Similarly to the operators defined in~\cite{Cousty2013} and~\cite{Dias2011}, the parameter $i$ controls how much of the complex will be affected by the operator. Figure~\ref{fig:OpClG} illustrates the operators $\opG[i]$ and $\clG[i]$ on two subcomplexes, depicted in gray. Since the dilation and erosion used to compose these operators affect less elements than the classical operators, we can expect the same behavior from them as well. 

It can be easily proven that the operators from definition~\ref{def:OpClG} act on complexes, that $\opG$ is an opening and $\clG$ is a closing. The families of operators formed by these two operators, indexed by the integer $i$, are a granulometry and an anti-granulometry, respectively. Therefore, they can be used to define new alternating sequential filters acting on $\C$. 

  \begin{defn}
    \label{def:asfG}
    Let $i\in\N$. We define:
    \begin{align}
      \forall X&\in\subcomplexes,\, \asfG\nolimits_{i}(X)=\nonumber\\
      & \paren{\opG[i]\clG[i]}  \paren{\opG[(i-1)]\clG[(i-1)]}\dots\paren{\opG[1]\clG[1]}(X)\\
      \forall X&\in\subcomplexes,\, \asfG[']\nolimits_{i}(X)=\nonumber\\
      & \paren{\clG[i]\opG[i]}  \paren{\clG[(i-1)]\opG[(i-1)]}\dots\paren{\clG[1]\opG[1]}(X)
    \end{align}
  \end{defn}

In this section we explored operators acting on subcomplexes composed by dimensional operators using a higher intermediary dimension. We defined an adjunction, families of openings and closings. We composed these granulometry and anti-granulometry into two alternating sequential filters. 

\subsection{Morphological operators on $\subcomplexes$ using a lower intermediary dimension}

We just explored compositions of dimensional operators using a higher intermediary dimension. We will explore compositions that use a lower intermediary dimension. As theorem~\ref{p:FamilyDownUp} suggests,  we can define a family of different operators, using the variation of the temporary dimension as parameter. However, we chose to explore only the operators that affects the smallest possible number of simplices, because such operators usually lead to more controlled filters. Additionally, one would need a space of higher dimensionality in order to properly exploit these families.

\begin{defn}
  Let $X\in\subcomplexes$. We define the operators $\dilS$ and $\eroS$ by:
  \label{def:DilEroS}
  \begin{align}
    \dilS& (X)=  \left\{ \bigcup\nolimits_{i\in[1\dots n]} \dilp{i-1}{i}\dilm{i}{i-1}(X) \right\} \bigcup\nonumber\\
    &\left\{\dilm{1}{0}  \dilp{0}{1}(X) \right\} \\
    \eroS&(X) = Cl\adjoint \left(  \left\{ \bigcup\nolimits_{i\in[1\dots n]} \erop{i-1}{i}\erom{i}{i-1}(X) \right\} \right.\nonumber\\
    &\bigcup\left\{\erom{1}{0}  \erop{0}{1}(X) \right\}          \bigg)
  \end{align}
\end{defn}

However, the following property states that the operators from definition~\ref{def:DilEroS} are the same operators from definition~\ref{def:dilG}.

\begin{proper}
  Let $i\in\N$ such that $1\leq i \leq (n-1)$.
  \begin{enumerate}
    \item{$\forall X\in\PC[i],\,\dilp{i-1}{i}\dilm{i}{i-1}(X)=\dilm{i+1}{i}\dilp{i}{i+1}(X)$;}
    \item{$\forall X\in\PC[i],\,\erop{i-1}{i}\erom{i}{i-1}(X)=\erom{i+1}{i}\erop{i}{i+1}(X)\!$.}
  \end{enumerate}
\end{proper}

This property can be proved by analysing the elements of the space that are included or removed by each operator. Following this property, the operators obtained using a lower intermediary dimension are identical to the ones obtained using a higher intermediary dimension. For this reason, we will only illustrate the results of one of them in the next section.

\section{Illustrations of some operators}
\label{c:exp}
We defined various operators and filters acting on subcomplexes. In this section we illustrate these operators, acting on values associated with elements of a mesh and on subcomplexes created from regular images. More results and quantitative comparisons are available in~\cite{Dias2012}. 

\subsection{Illustration on a tridimensional mesh}
\label{sec:mesh}
\begin{figure}
  \centering 
  \subfigure[$X$ (in black).]  {
    \includegraphics[width=0.12\textwidth]{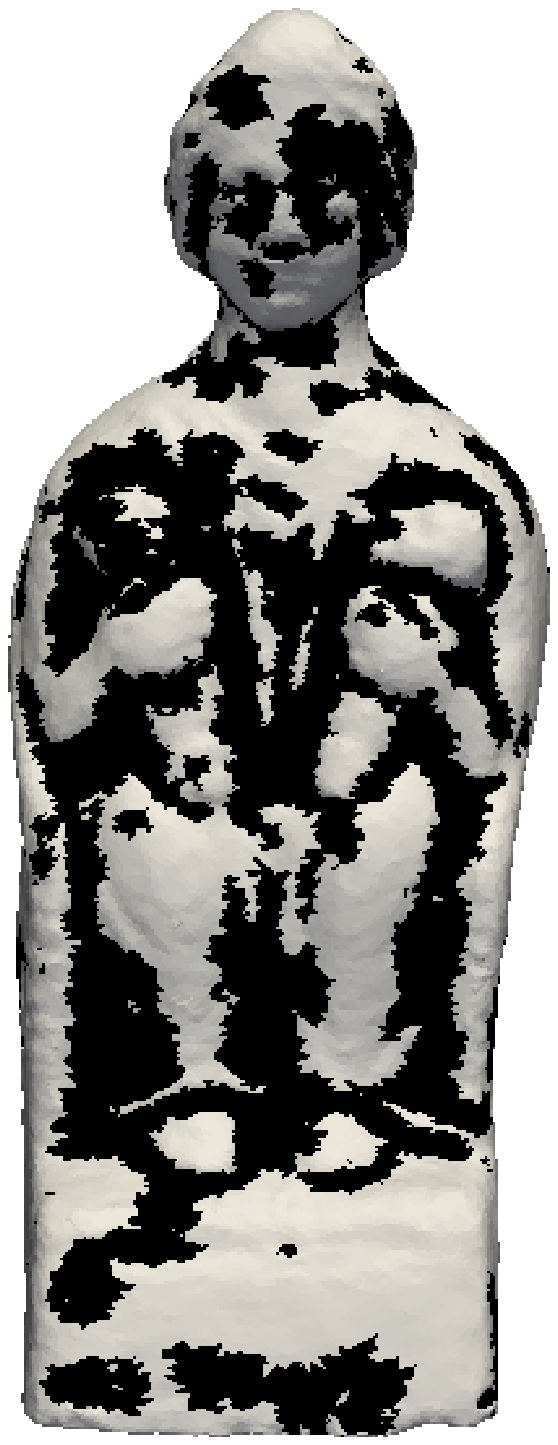}
    \label{fig:StThres}  } 
  \subfigure[${\asfG_{3}}(X)$.]  {
    \includegraphics[width=0.12\textwidth]{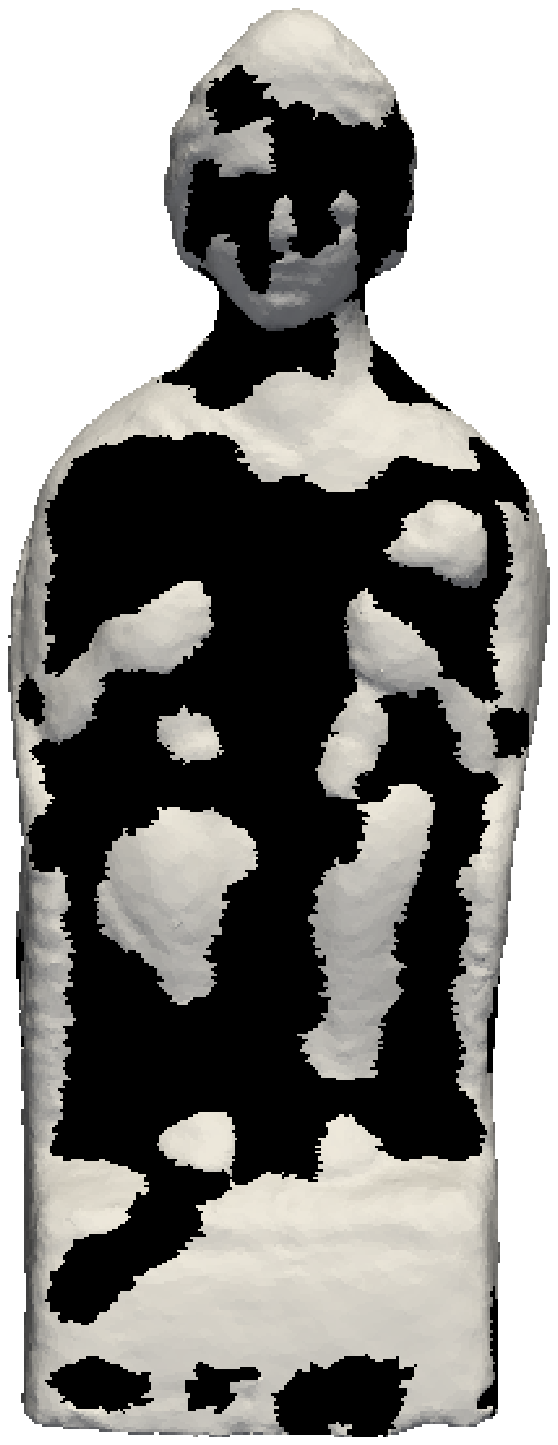}
    \label{fig:StG}  }
  \subfigure[${\asfG[']_{3}}(X)$.]  {
    \includegraphics[width=0.12\textwidth]{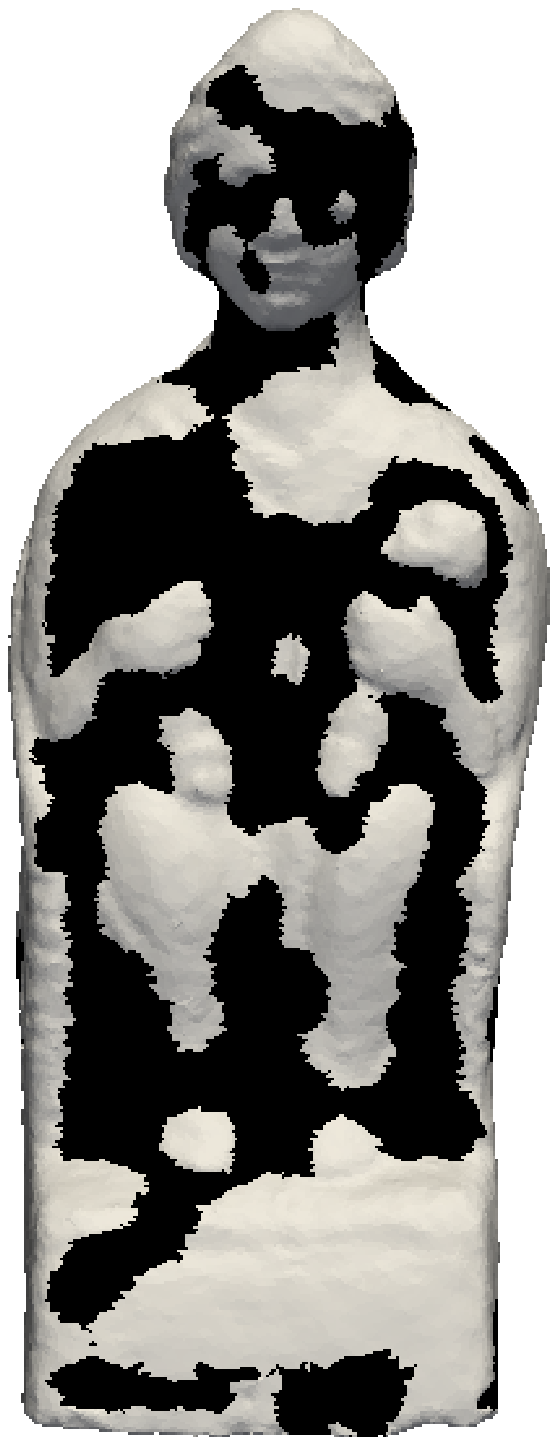}
    \label{fig:StGl}  } 
  \caption{Rendering of the mesh considered. The sets are represented in black. [see text] \label{fig:exp:stC}\label{fig:exp:stG}}
\end{figure}

As illustration, we processed the curvature values associated with a $3$D mesh, courtesy of the French Museum Center for Research. We computed the curvature for the vertices and propagated these values to the edges and triangles, following the procedure described in~\cite{Alcoverro2008}, resulting in values between $0$ and $1$. These values were then processed using our filters. For visualization purposes only, we thresholded the values at $0.51$, as shown in black on figure~\ref{fig:StThres} that depicts the thresholded set for the original curvature data. The renderings presented in this section consider only the values associated with the vertices of the mesh, and no interpolation was used.

\subsection{Illustration on binary regular images}
\label{sec:IlusIm}
In this section we consider the application of our alternating sequential filters on regular images. For this end, we need to create a simplicial complex based on the image. Several methods can be used and the choice is application dependent. Here, we create a vertex for each pixel, with edges between the vertices, six for each vertex, corresponding to an hexagonal grid. Triangles are placed between three vertices, so each vertex belongs to six triangles. We then consider the greatest complex that can be made using the value of the vertices. For visualization purposes, the images presented depict only the values associated with the vertices.

\begin{figure}
  \centering 
  \subfigure[Original image.]  {
    \includegraphics[width=0.21\textwidth]{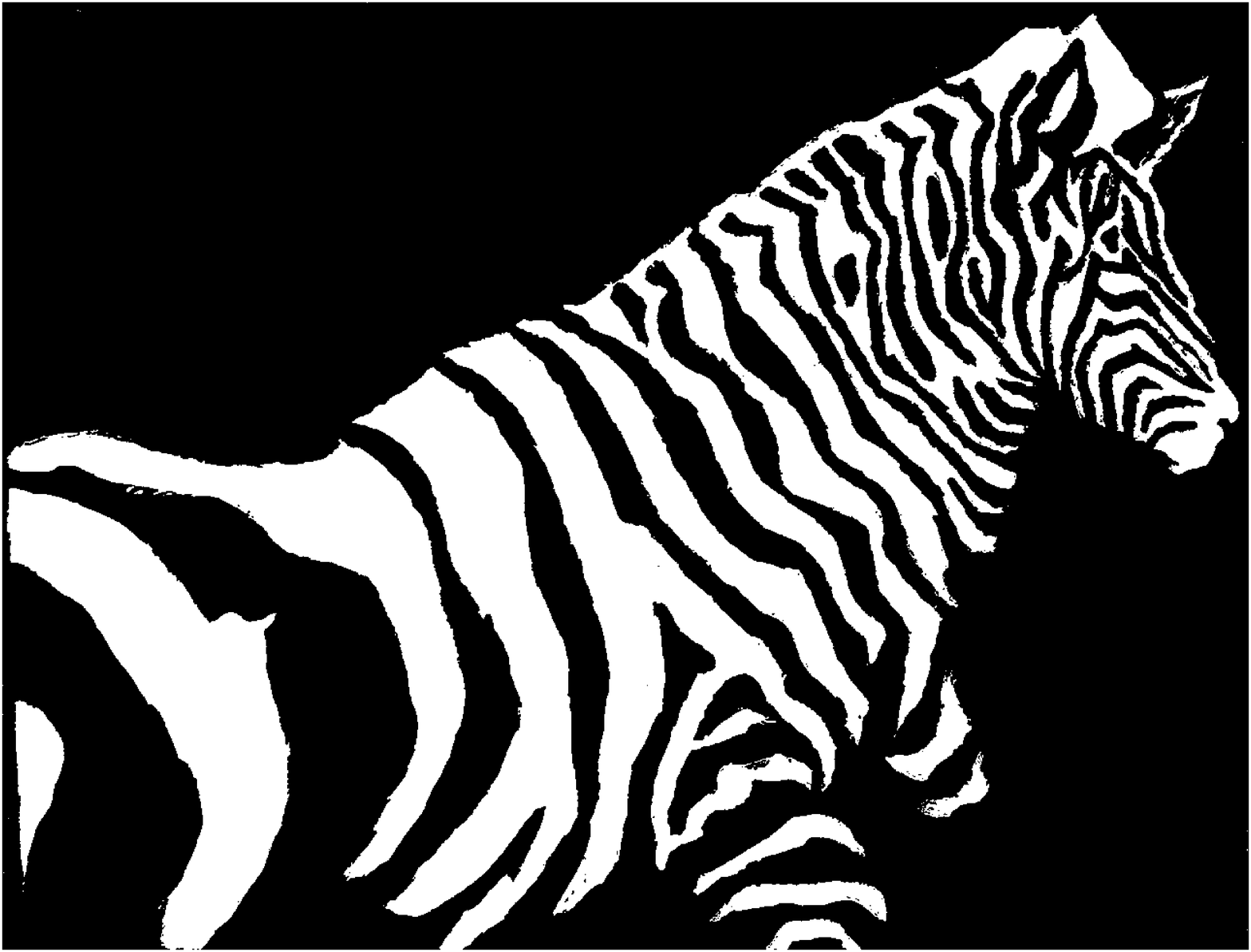}
    \label{fig:ZebOrig}  
		} 
  \subfigure[Noisy version.]  { 
    \includegraphics[width=0.21\textwidth]{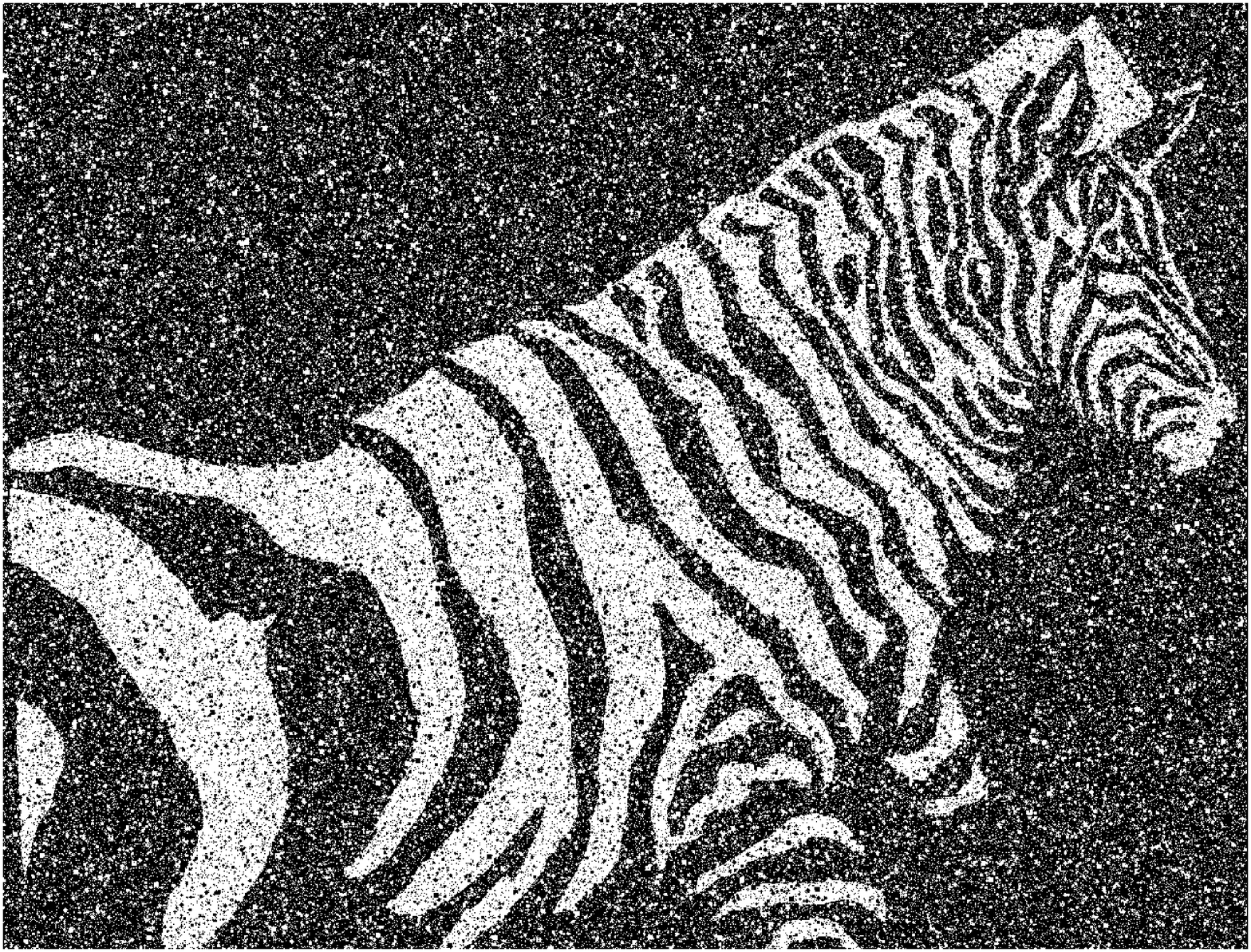}
    \label{fig:ZebNoisy}  
		} 
  \subfigure[$\asfG_{6}$.]{ 
   \includegraphics[width=0.21\textwidth]{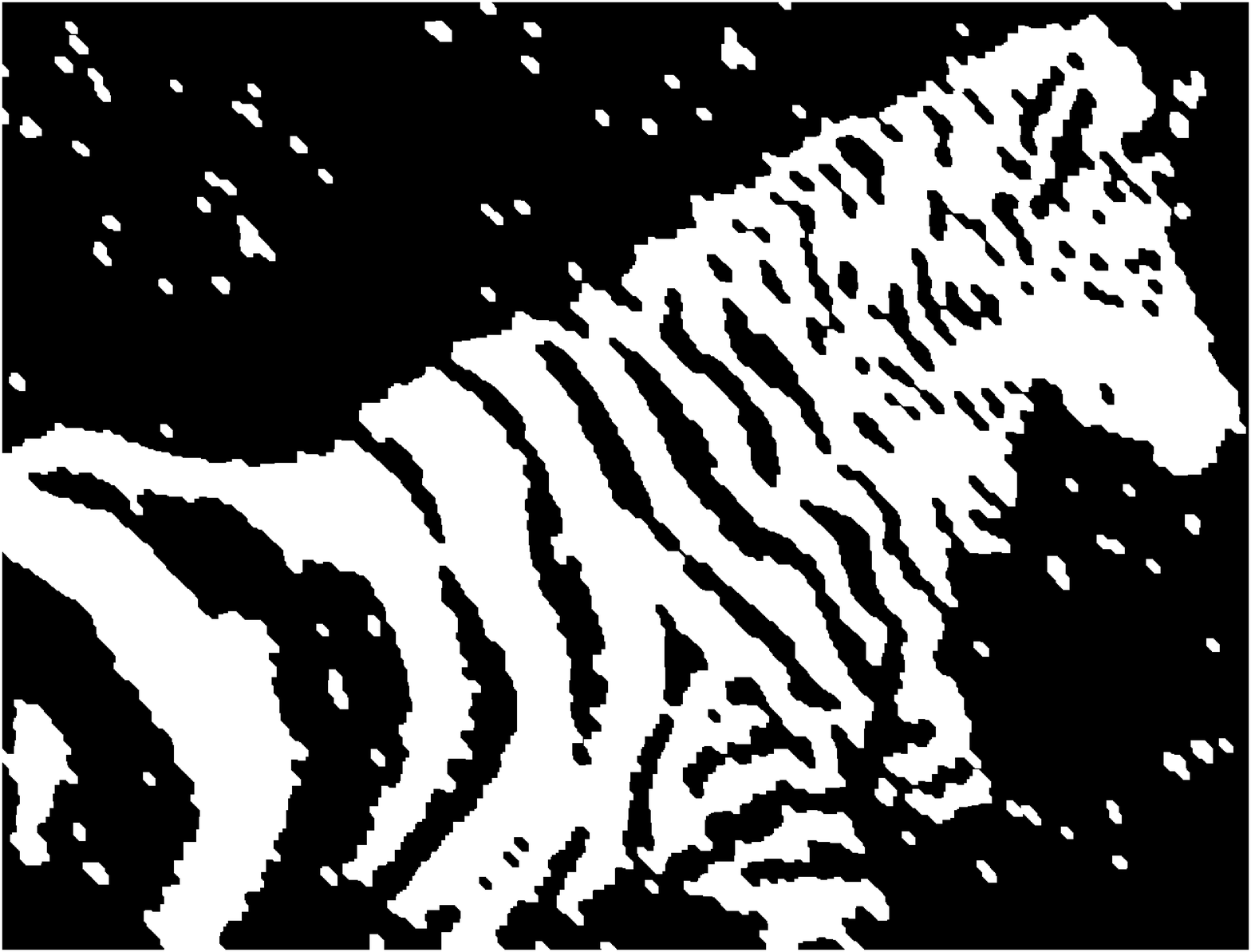}
    \label{fig:ZebASFG}  
		} 
  \subfigure[${\asfG}'_{3}$.]{ 
   \includegraphics[width=0.21\textwidth]{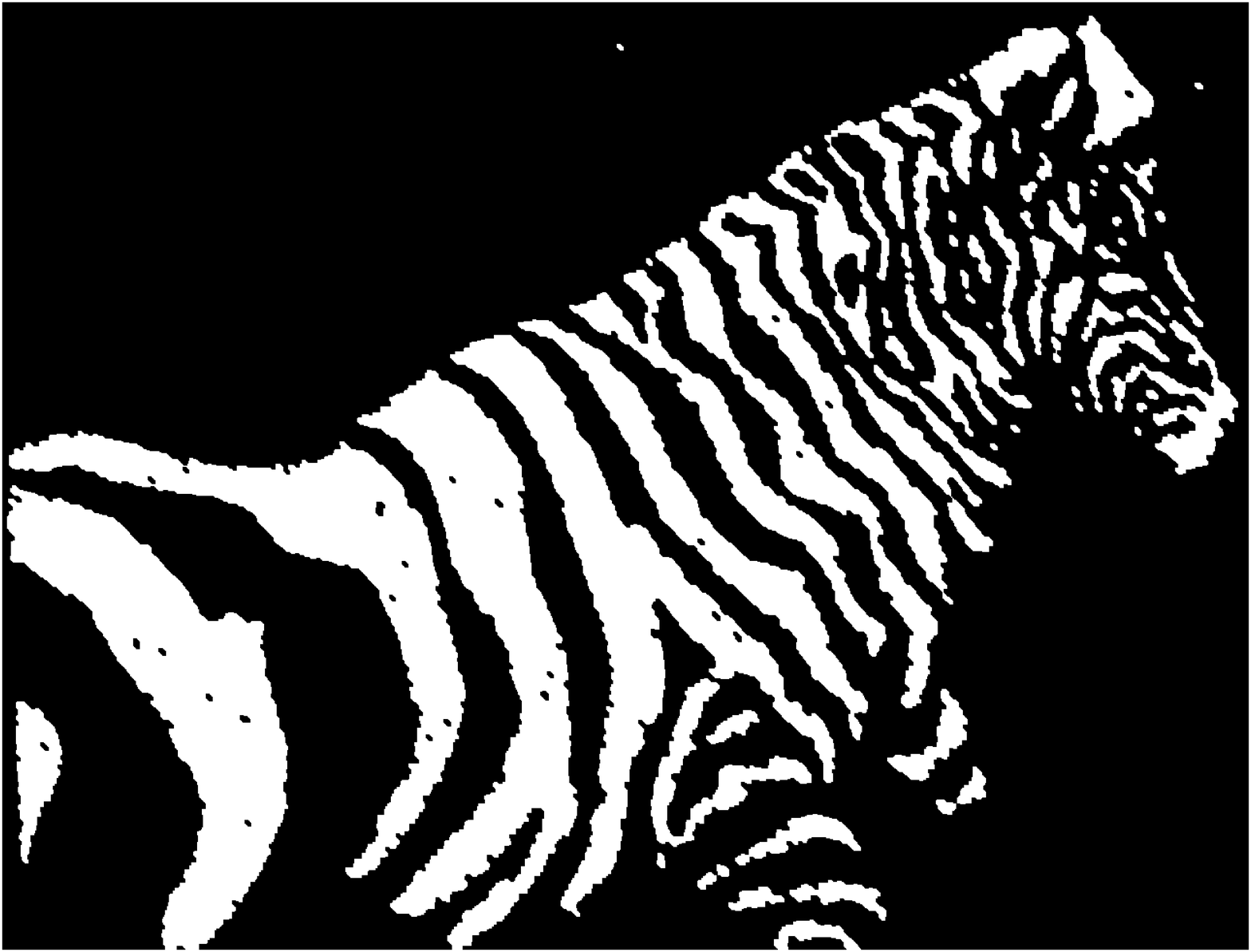}
    \label{fig:ZebASFGl}  
		} 
	\subfigure[Graph $ASF_{6/2}$.]{ 
   \includegraphics[width=0.21\textwidth]{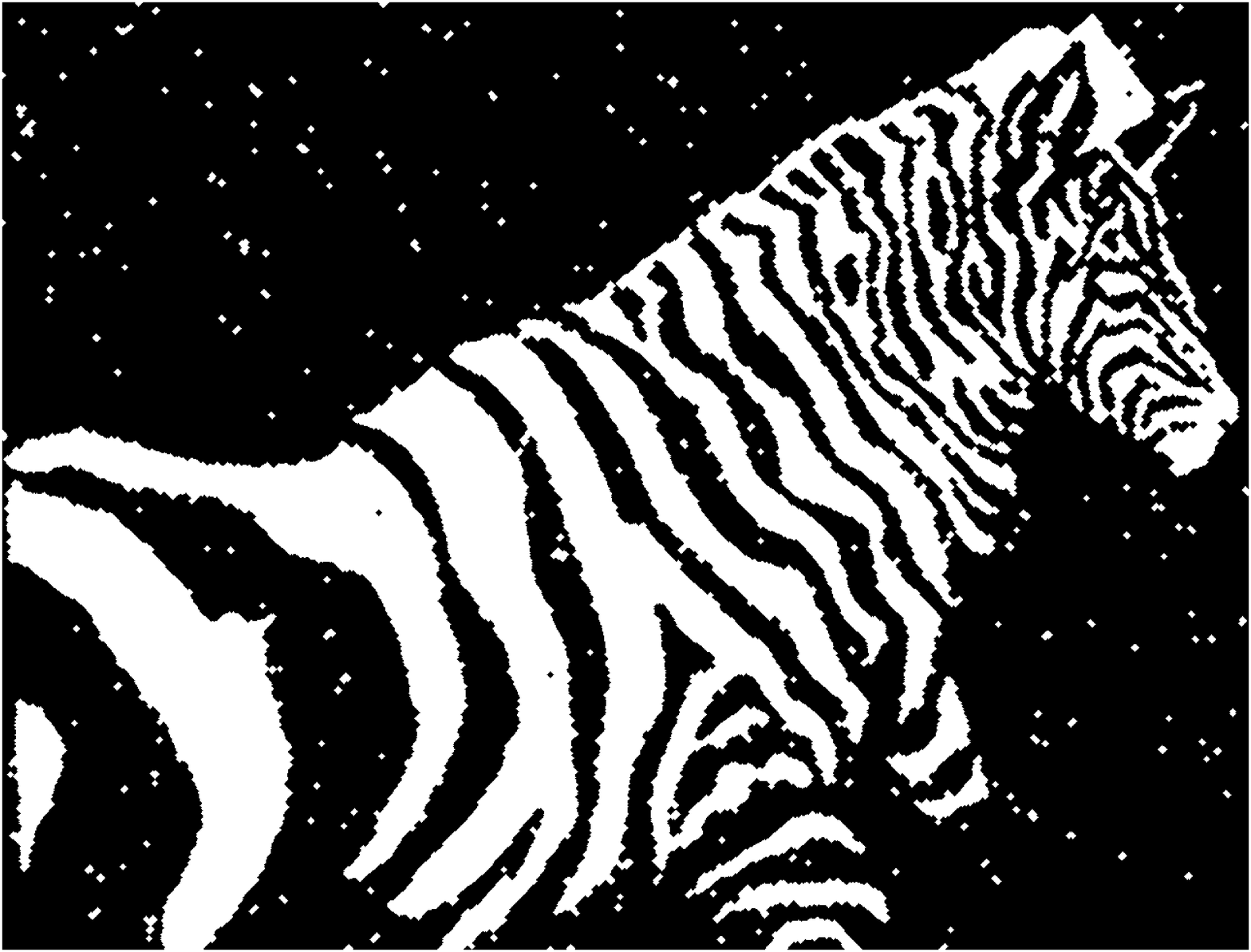}
    \label{fig:ZebGrASF6}
  } 
	\subfigure[$\asfC_{3}$.]{ 
   \includegraphics[width=0.21\textwidth]{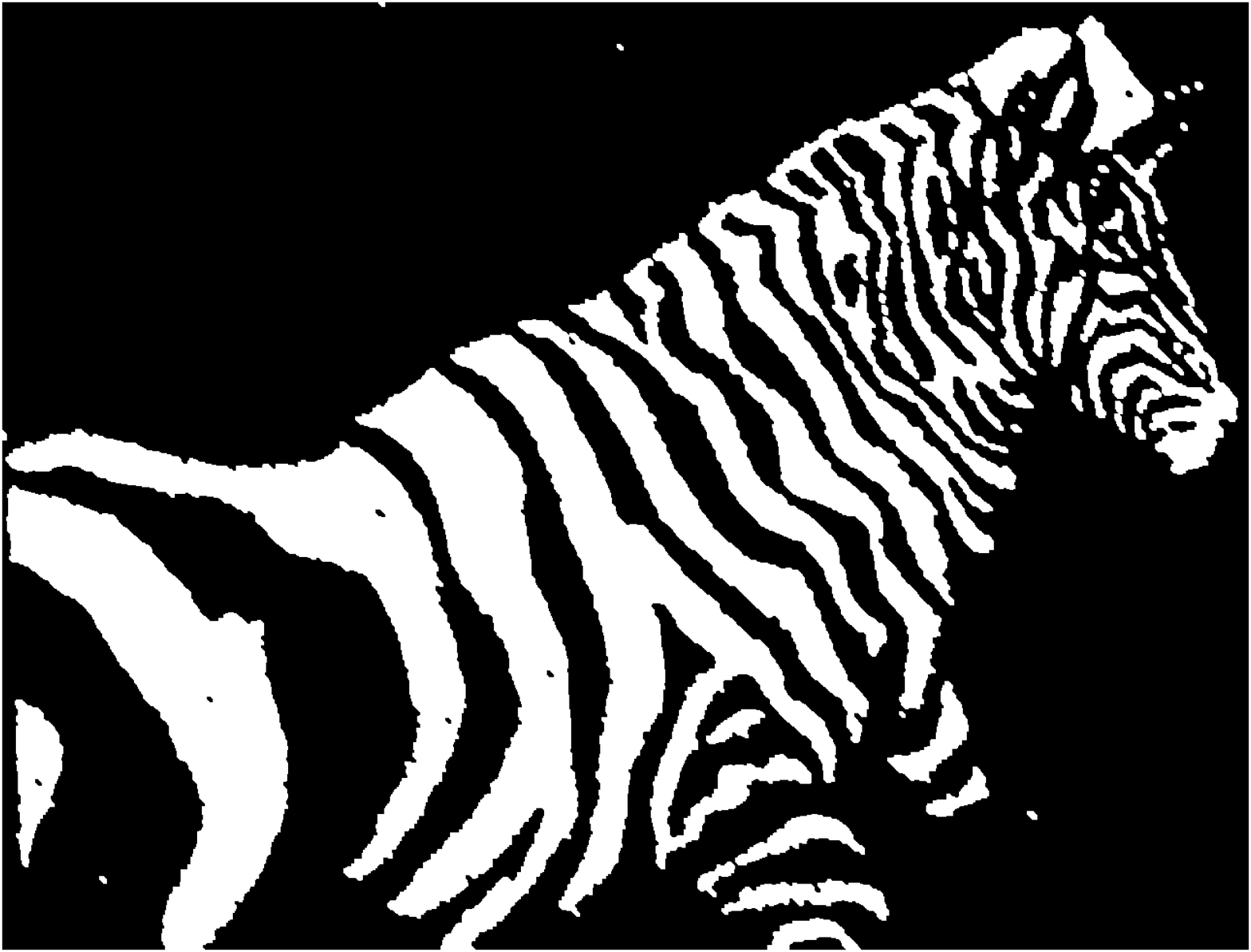}
  }
  \caption{Illustration of some results obtained with the operators $\asfG$ and related literature results. Images $(a)$, $(b)$ and $(e)$ are from~\cite{Cousty2009} and image $(f)$ is from~\cite{Dias2012}. \label{fig:ZebIllG}}
\end{figure}

We compare our results with the literature considering the same image used by Cousty~\etal~\citep{Cousty2009}, shown on figure~\ref{fig:ZebOrig}. The noisy image shown on figure~\ref{fig:ZebNoisy} was processed. Figure~\ref{fig:ZebIllG} shows some results of the operators $\asfG$ and $\asfG[']$, along with some results from~\cite{Cousty2013} and~\cite{Dias2012}, for visual comparison. The operator $\asfG$ removed most of the features of the zebra and left some noise on the background. The operator $\asfG[']$ removed most of the background noise, while preserving some of the gaps between the stripes. However it also removed the smaller features of the object and left small holes. From these results, we may conclude that our operators are, for this type of image, on a competitive level with the operators presented in the literature. 

Additionally, Mennillo~\etal~\citep{Mennillo} used the dimensional operators for document processing as a pre-processing stage to boost OCR performance with encouraging results.

While the results for regular image processing are good, they do not fully exploit the structure of the simplicial complex, nor the flexibility of the operators. We expect the filtering results to be even better when considering more complex scenarios.

\subsection{Illustration on a grayscale image}
\begin{figure}
  \centering 
  \subfigure[Original.]{
    \includegraphics[width=0.20\textwidth]{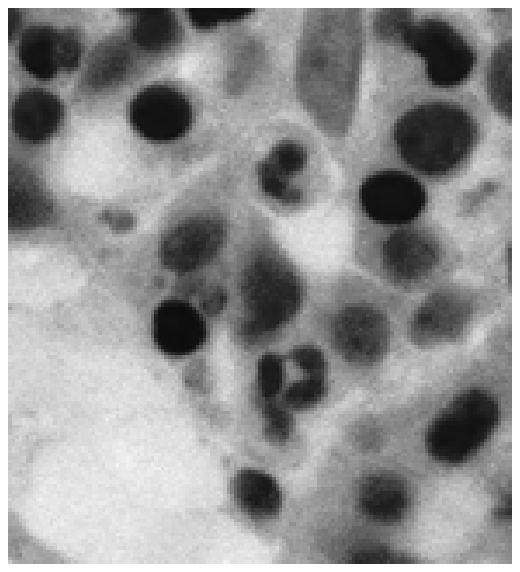}  
    \label{fig:ExGray}
  }
  \subfigure[Classical closing.]{ 
    \includegraphics[width=0.20\textwidth]{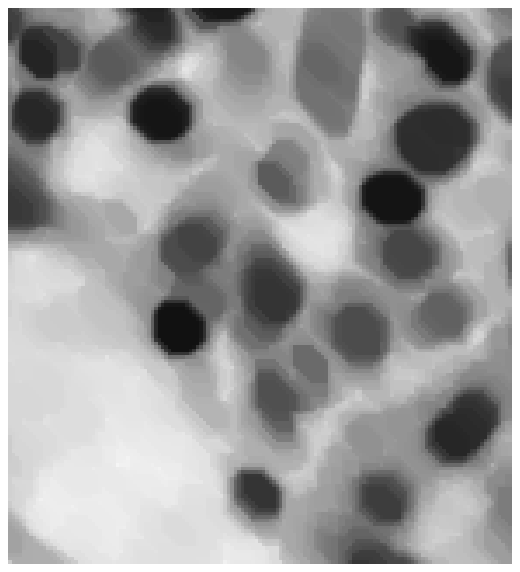}  
  } 
  \subfigure[$\phi^{cm}_{12/3}$]{ 
    \includegraphics[width=0.20\textwidth]{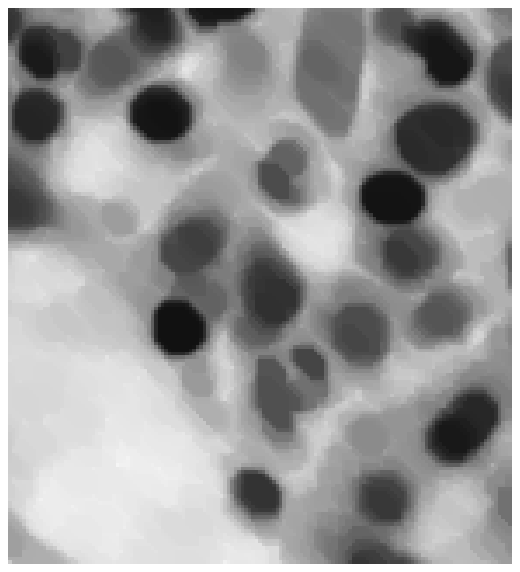}  
  } 
  \subfigure[${\clG}_4$]{ 
    \includegraphics[width=0.20\textwidth]{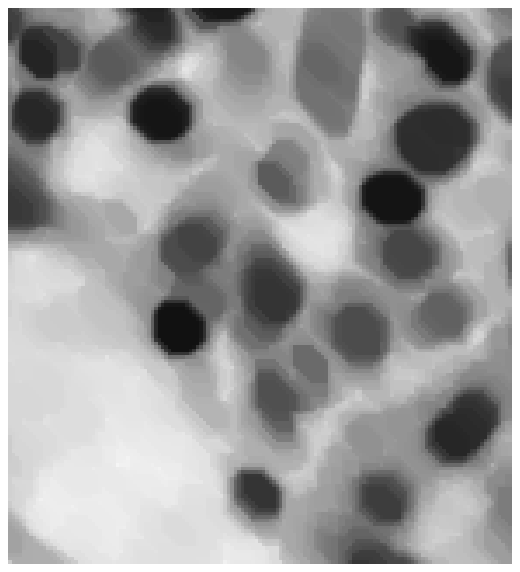}  
  } 
  \caption{Zoom of the same section of the image after closings of size $4$.\label{fig:MicroCl}}
\end{figure}

We now consider a grayscale image, with a small part shown on figure~\ref{fig:ExGray}. This image is from Jon Salisbury at the english language wikipedia, released under Creative Commons license and is a photomicrograph of bone marrow showing abnormal mononuclear megakaryocytes, typical of $5q-$ syndrome. The image was converted from RGB to grayscale. Medical meaning aside, this image was chosen because it has many features of various sizes.

Figure~\ref{fig:MicroCl} shows the results of the considered closing operators, with size $4$. As expected, all operators removed the small noise of the image. The result of the operator $\clG[4]$ was identical to the classical operator, because we only depict the values of the vertices, while the operator $\phi^{cm}_{12/3}$ was generally less abrasive, maintaining the level of deeper holes that were raised by the classical operator.


\section{Conclusion}
\label{ch:conclusion}
In this work we explored the dimensional operators, presenting composition properties and defining new operators. We created new dilations, erosions, openings, closings and alternating sequential filters. We also used these operators to express operators from the related literature, acting on digital objects, such as graphs and simplicial complexes.

\section{Acknowledgments}

This work has been funded, in part, by Conseil Général de Seine-Saint-Denis and french Ministry of Finances through the FUI6 project "Demat-Factory". Fábio Dias received additional funding by FAPESP $2012/15811-1$. Jean Cousty and Laurent Najman received funding from the Agence Nationale de la Recherche, contract $ANR-2010-BLAN-0205-03$.

\bibliographystyle{elsarticle-harv}
\bibliography{diasAMM}

\end{document}